\documentclass[aps, twocolumn, prl, citeautoscript, superscriptaddress, amsmath, 8pt]{revtex4-2}

\usepackage{graphicx}
\usepackage{gensymb}
\usepackage{color}
\usepackage[dvipsnames]{xcolor}
\usepackage{float}
\usepackage{upgreek}
\usepackage{bm}
\usepackage{amsmath,amssymb}
\usepackage[colorlinks=true,citecolor=blue,linkcolor=blue]{hyperref}
\setcitestyle{super}

\begin{document}

\title{Spin-dependent photoluminescence in carbon-based quantum dots}

\author{Erin S. Grant}
\thanks{These authors contributed equally}
\affiliation{Department of Physics, School of Science, RMIT University, Melbourne, VIC 3001, Australia}

\author{Joseph F. Olorunyomi}
\thanks{These authors contributed equally}
\affiliation{Applied Chemistry and Environmental Science, School of Science, RMIT University, Melbourne, VIC 3001, Australia}
\affiliation{Manufacturing Research Unit, CSIRO, Clayton, VIC 3168, Australia}

\author{Sam C. Scholten}
\affiliation{Department of Physics, School of Science, RMIT University, Melbourne, VIC 3001, Australia}

\author{Islay O. Robertson}
\affiliation{Department of Physics, School of Science, RMIT University, Melbourne, VIC 3001, Australia}

\author{Amanda N. Abraham}
\affiliation{Department of Physics, School of Science, RMIT University, Melbourne, VIC 3001, Australia}

\author{Nandish H. Srikantamurthy}
\affiliation{Applied Chemistry and Environmental Science, School of Science, RMIT University, Melbourne, VIC 3001, Australia}
\affiliation{Manufacturing Research Unit, CSIRO, Clayton, VIC 3168, Australia}

\author{Billy J. Murdoch}
\affiliation{RMIT Microscopy and Microanalysis Facility, STEM College, RMIT University, Melbourne, VIC 3001 Australia}

\author{Edwin L. H. Maye}
\affiliation{School of Science, RMIT University, Melbourne, VIC 3001, Australia}

\author{Blanca del Rosal Rabes}
\affiliation{Department of Physics, School of Science, RMIT University, Melbourne, VIC 3001, Australia}

\author{Alexander J. Healey} 
\affiliation{Department of Physics, School of Science, RMIT University, Melbourne, VIC 3001, Australia}

\author{Cara M. Doherty}
\affiliation{Manufacturing Research Unit, CSIRO, Clayton, VIC 3168, Australia}

\author{Philipp Reineck}
\affiliation{Department of Physics, School of Science, RMIT University, Melbourne, VIC 3001, Australia}

\author{Xavier Mulet}
\email{xavier.mulet2@rmit.edu.au}
\affiliation{Applied Chemistry and Environmental Science, School of Science, RMIT University, Melbourne, VIC 3001, Australia}

\author{Jean-Philippe Tetienne}
\email{jean-philippe.tetienne@rmit.edu.au}
\affiliation{Department of Physics, School of Science, RMIT University, Melbourne, VIC 3001, Australia}

\author{David A. Broadway}
\email{david.broadway@rmit.edu.au}
\affiliation{Department of Physics, School of Science, RMIT University, Melbourne, VIC 3001, Australia}

\begin{abstract}
The ability to modulate the photoluminescence (PL) of nanomaterials via spin-related effects is vital for many emerging quantum technologies, with nanoscale quantum sensing and imaging being particular areas of focus. Carbon-based quantum dots (CQDs) are among the most common forms of luminescent nanomaterials, appealing due to their ease of synthesis, tunability through organic chemistry, high brightness, and natural biocompatibility. However, the observation of room-temperature, spin-dependent PL has remained elusive. Here we report on the observation of PL modulation of CQDs by magnetic fields  ($\sim 10$\,mT) under ambient conditions. We synthesize a series of CQDs using 19 different amino acids, which have a range of PL emission spectra and exhibit a clear magneto-PL effect (up to $\sim1\%$ change). Furthermore, an electron spin resonance is detected in the PL with a $g$-factor of $g\approx2$, suggesting a process similar to the radical pair mechanism is responsible. Finally, we show that the magneto-PL contrast decreases in the presence of paramagnetic species, which we attribute to an increase in magnetic noise-induced spin relaxation in the CQDs. Our work brings new functionalities to these commonly used and biocompatible luminescent nanoparticles, opening new opportunities for \textit{in situ} quantum sensing and imaging of biological samples.
\end{abstract}

\maketitle
\section*{\label{Intro}Introduction}
Nanomaterials that enable nanoscale quantum sensing or imaging via optically addressable spins have steadily been gaining attention over the last decade~\cite{alfieri_nanomaterials_2023, aslamQuantumSensorsBiomedical2023, wolfowiczQuantumGuidelinesSolidstate2021}.
There are a number of material classes which host such properties, each with pros and cons. So far, these include spin defects in wide band gap materials like diamond~\cite{zhang_toward_2021, casolaProbingCondensedMatter2018, wangFluorescentNanodiamondsNanoscale2025}, hBN~\cite{vaidya_quantum_2023, scholten_multi-species_2024}, SiC~\cite{castelletto_quantum_2023}, spin-active fluorescent proteins~\cite{evans_magnetic_2013,xiang_mechanism_2025, feder_protein_2025, abrahams_quantum_2024}, and other molecules~\cite{mena_room-temperature_2024, wang_giant_2024, koppOpticallyDetectedCoherent2025, chowdhuryBrightTripletBright2025}. 
Of these, diamond and SiC nanoparticles -- crystalline materials that host lattice defects with an associated spin -- are the most well-studied~\cite{robertsQuantumSensingSpin2025, suFluorescentNanodiamondsQuantum2025}. 
Although useful for sensing, with comparatively long coherence times, they are generally large ($\gg 20$\,nm)~\cite{torelli_perspective_2019}. This limits their utility for intracellular measurements, where particles below 10 nm are desirable. 2D materials like hBN offer the potential to create much smaller sensors, but lateral particle dimensions are generally still large, on the nanometer scale, and remain undeveloped as colloidal biomarkers in biology~\cite{robertson_detection_2023, gaoQuantumSensingParamagnetic2023}. An emerging class of optical-spin sensors are the magneto-sensitive fluorescent proteins, which have seen rapid development over the last few years~\cite{xiang_mechanism_2025, feder_protein_2025, abrahams_quantum_2024}. However, they require genetic encoding and expression in cells, which is a lengthy process that can limit their applications. Similarly, a number of molecular-based spin sensors have been identified, but these often require cryogenic temperatures and have not been considered in the context of biological systems~\cite{mena_room-temperature_2024, wang_giant_2024,koppOpticallyDetectedCoherent2025, chowdhuryBrightTripletBright2025}.

Another well-studied nanomaterial is the carbon-based quantum dot (CQD), a broad class of materials generally derived from molecular precursors that has garnered interest for its bright photoluminescence (PL) and biocompatibility~\cite{guan_emerging_2023, gidwani_quantum_2021, wang_light_2022, sun_recent_2023}. In addition to these favourable qualities, CQDs are generally small (\textless 10 nm) and can be synthesized cost-effectively at scale~\cite{manikandan_green_2022}. Combined, these attributes make CQDs an attractive material class for spin-based sensing which could facilitate widespread adoption of quantum sensing techniques in chemistry and biology. However, to date spin-dependent PL has only been reported under cryogenic temperatures and ultra-high magnetic fields~\cite{chen_magnetic_2016, zhang_magnetic_2024}.

Here we report, for the first time, room-temperature magneto-PL and optically detected electron spin resonance in CQDs. Specifically, we perform PL measurements on a series of CQD samples derived from 19 different amino acids (\textit{aa}-CQDs) and observe that the vast majority (16 out of 19) of the samples tested exhibit magnetic-field dependent PL (MPL) at relatively low magnetic fields ($\sim 10$\,mT) and at room temperature, both in a dry phase and in suspension. Furthermore, by applying radiofrequency (RF) fields we can directly observe an electron spin resonance (ESR), revealing a $g$-factor of $g\approx2$. Our combined MPL and ESR observations are consistent with a radical pair like mechanism \cite{evans_magnetic_2013}, whereby a photo-generated weakly coupled spin pair experiences singlet-triplet spin mixing which is influenced both by static magnetic fields and resonant RF fields. Finally, we show that the MPL contrast is modulated by the presence of paramagnetic species, similarly to other spin-based optical defects \cite{Steinert2013, Ermakova2013, Kaufmann2013,robertson_detection_2023,gaoQuantumSensingParamagnetic2023}, even when no significant change in the PL spectrum (a commonly used sensing indicator~\cite{guan_emerging_2023, gidwani_quantum_2021, wang_light_2022}) is observed. 
These results suggest that CQDs could form a broad class of easy-to-make optical-spin sensors, with significant potential for future optimisation through chemical engineering. 

\section{\label{Results}Results and Discussion}

\begin{figure}
    \centering
    \includegraphics[width=\columnwidth]{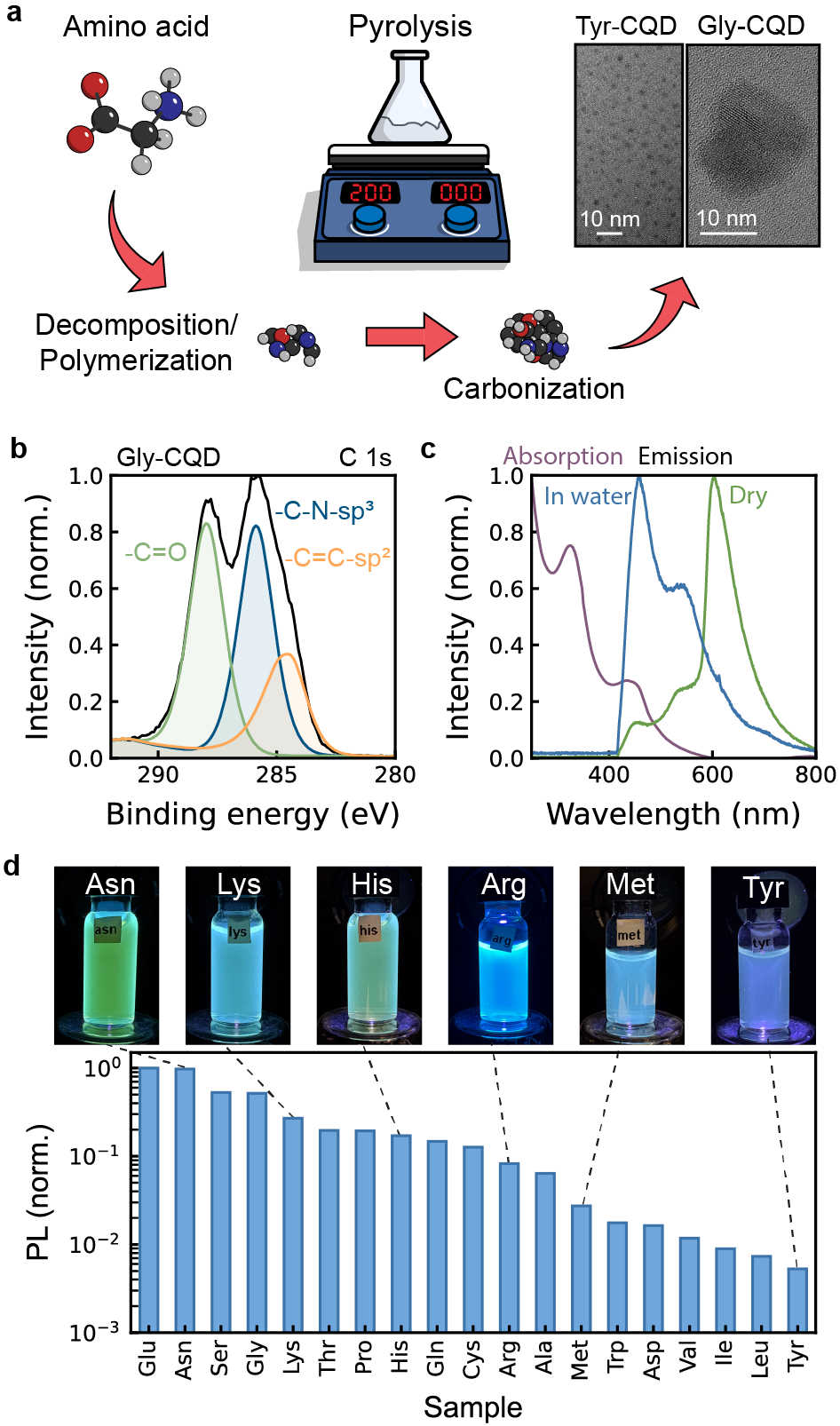}
    \caption{\textbf{Synthesis of amino acid-derived carbon-based quantum dots.}
    \textbf{a} Schematic showing direct conversion of amino acids to carbon-based quantum dots (CQD) via pyrolysis, and TEM images of two different amino acid dervied \textit{aa}-CQDs. 
   \textbf{b} High resolution C 1s XPS spectrum of the \textit{Gly}-CQD sample. 
    \textbf{c} Absorption (purple trace) and PL emission spectrum (blue trace) of \textit{Gly}-CQDs in water, and PL spectrum of the same particles dried on glass (green trace). PL spectra were acquired using 405\,nm excitation.
    \textbf{d} Photos of a few samples under 365\,nm illumination. The bar graph indicates the photoluminescence intensity of all 19 samples (normalised to the brightest) in water under 405\,nm excitation, collected through a 420~nm longpass filter. The samples are ordered by decreasing brightness. 
    }
    \label{fig1}
\end{figure}

The \textit{aa}-CQDs were synthesized via pyrolysis of amino acids.  This process involves the sequential transformation processes of molecular decomposition (accompanied by the release of volatile condensates), polymerisation, carbonisation, and surface passivation (Fig.~\ref{fig1}\textbf{a}), followed by extraction into water. This approach has merits over more complex multi-step hydrothermal synthesis~\cite{kumar_high_2024}, due to its simplicity, fixed chemical composition (primarily determined by the amino acid used) which avoids solvent influence, and scalability.
The pyrolysis begins by placing a crystalline amino acid powder in an open glass flask and heat it at the respective melting point, ranging from 200°C to 340°C, under ambient atmospheric conditions. 
After this, the \textit{aa}-CQDs are suspended in a solution phase by adding Milli-Q water to the carbonized residue and vigorously stirring, resulting in a stable colloid solution. Following this, the sample is subjected to several freeze-thaw purification cycles to produce a final suspended sample (see Methods for details). Because amino acids are thermally unstable,~\cite{Weiss2018} this simple pyrolysis enables the rapid synthesis of new materials, with their unique properties determined by the chemical structures of the precursors.

\begin{figure*}
    \centering
    \includegraphics{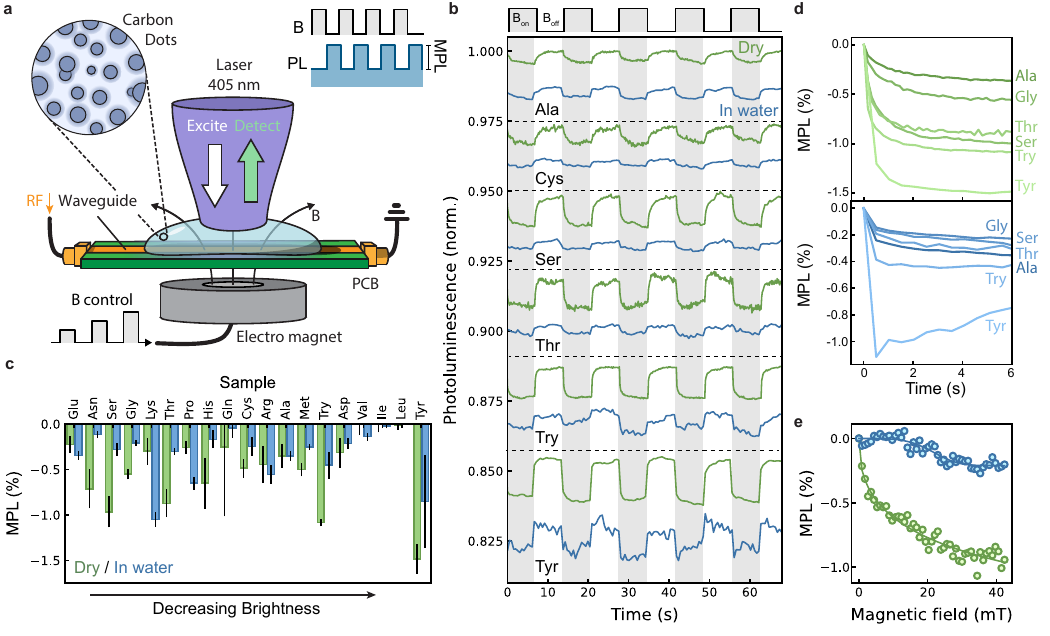}
    \caption{\textbf{Magneto-photoluminescence (MPL) in \textit{aa}-CQDs.}
    \textbf{a} Schematic of the experiment, where \textit{aa}-CQDs are placed on a printed circuit board (PCB) for RF control (used in Fig.~\ref{fig3}), with an electromagnet positioned underneath for $B$-field control. The light excitation and collection is performed from above and separated with appropriate filters. 
    \textbf{b} PL response to switching magnetic fields on and off for a few exemplar samples both dry (green) and in water(blue). The sampling rate was varied between 2 and 6.25 Hz depending on brightness, to obtain an acceptable signal-to-noise ratio.
    \textbf{c} Histogram of the MPL contrast observed for the 19 \textit{aa}-CQDs samples dry (green) and in water (blue) form. The error bar is the standard deviation from multiple on-off cycles. MPL is confirmed in all but 3 samples for which the error bar is larger than the mean.
    \textbf{d} Example time traces of the PL when the magnetic field is turned on for a few dry (top panel, green) and in-water (bottom panel, blue) samples. Data for all samples can be found in the SI (Fig. S13).
    \textbf{e} Maximum MPL contrast as a function of the applied magnetic field for \textit{Gly}-CQDs dry (green) and in water (blue) form. Solid lines are stretched-exponential fits.
    }
    \label{fig2}
\end{figure*}

Interestingly, several of the \textit{aa}-CQD samples are composed of crystalline cores (determined by powder x-ray diffraction (XRD) and transmission electron microscopy (TEM), see SI) with functional groups that passivate the surface.
For example, \textit{Gly}-CQD particles (derived from Glycine, Gly) were seen to have crystalline core (see TEM images in Fig.~\ref{fig1}\textbf{a} and additional characterisation in SI). 
This crystalline behaviour is uncommon in CQDs and more closely aligns with graphene and graphitic quantum dots, which typically have similarly large sizes ($\approx$20 nm) and display sharp XRD peaks~\cite{kim_anomalous_2012, niu_graphene-like_2012}.
However, we observed a large distribution of both sizes and crystallinities between the different \textit{aa}-CQD samples studied. For example, \textit{Tyr}-CQDs are significantly smaller ($\sim 2~$nm) and more amorphous compared to \textit{Gly}-CQDs (see SI for more examples).
The high-resolution C 1s XPS spectrum provides information about the functional groups present on the surface of the \textit{aa}-CQDs. Fig.~\ref{fig1}\textbf{b} shows an example spectrum from \textit{Gly}-CQD, revealing characteristic features consistent with surface functionalisation by oxygen- and nitrogen-containing groups. A peak centred at approximately 288.0 eV corresponds to the O=C– signal, indicative of carbonyl or carboxyl functional groups. The main hydrocarbon peak, located at around 286 eV, displays considerable broadening, which upon fitting, is attributed to distinct contributions from sp$^3$-hybridised –C–N– species ($\sim285.9$ eV) and sp$^2$-hybridised –C=C– ($\sim 284.5$ eV). See SI Fig. S3 for further spectra. Compared to their amino acid precursors, the \textit{aa}-CQDs have an increased XPS signal from atomic C and reduced O, N, and S, indicating dehydration and volatilisation of heteroatom-containing species (Fig. S1). 
FTIR fingerprinting was performed on different batches of samples synthesised at various times to demonstrate the reproducibility of our approach. 
The key feature is the appearance of a peak at 1595-1520 cm${-1}$, which demonstrates the formation of unsaturation in the CQDs (Fig. S2).

The optical absorption and PL spectra of the \textit{aa}-CQDs vary greatly with respect to the specific amino-acid precursor, as well as between the dry phase and suspension. As an example (Fig.~\ref{fig1}\textbf{c}), in water the \textit{Gly}-CQDs absorb mainly in the UV and blue (200-500\,nm) and emit in the blue-green region under 405\,nm illumination. However, when dried onto a substrate the spectrum red shifts (spectral data for other samples are shown in the SI Fig. S6 and S9). 
In Fig.~\ref{fig1}\textbf{d}, we plot the PL emission intensity of all 19 samples in water under the same illumination conditions (405\,nm, 50~mW). The dimmest sample (\textit{Tyr}-CQD) is over two orders of magnitude dimmer compared to the brightest (\textit{Glu}-CQD). The inset images in Fig.~\ref{fig1}\textbf{d} show samples under UV (365\,nm) illumination, where the differences in peak emission wavelength are evident by eye. 

To explore the magneto-optical properties of the \textit{aa}-CQDs, we drop cast CQDs suspended in water onto glass substrates (see Methods for details) and place them on a PCB patterned with an RF waveguide (which will be used later to drive the spin resonance) positioned above an electromagnet to control the applied magnetic field (Fig.~\ref{fig2}\textbf{a}). 
The samples are excited using a 405~nm laser, and the PL is filtered with a 405~nm longpass filter and collected by a camera. 
The magneto-photoluminescence (MPL) effect is characterised by repetitively switching the magnetic field between $B = 0$ and $46$~mT and observing the induced PL change. Example time traces are shown in Fig.~\ref{fig2}\textbf{b} for a few samples, where the $B$-field is switched every 7 seconds. Note that the background bleaching, typical for CQDs~\cite{dua_stability_2023}, has been subtracted, see SI Fig. S12. The MPL contrast, defined as the relative PL change at the end of the 7\,s period, $C_M=[{\rm PL}(B)-{\rm PL}(0)]/{\rm PL}(0)$, is always negative, i.e. the PL is reduced when a magnetic field is applied. The contrast is clearly measurable in the majority of samples (16 out of 19 samples have a contrast exceeding the statistical uncertainty) and reaches up to -1.5\%, see Fig.~\ref{fig2}\textbf{c}. The effect is generally larger in the dry phase, \textit{Lys}-CQD being an exception with -1.0\% in water compared to -0.3\% in the dry phase. There is no correlation between MPL contrast and overall brightness.   

The time dynamics of the MPL response also vary greatly as illustrated by the example decay traces (i.e. PL after the $B$-field is turned on) in Fig.~\ref{fig2}\textbf{d}. We generally observe a multi-component decay, with a fast component evidenced by an initial step (indicating a characteristic time of the same order or below our time resolution of $\sim100$~ms) and a slower component which takes several seconds to settle. In some cases, the slow component has an opposite sign to the fast component i.e. PL is partially recovered after the initial drop (see response of \textit{Tyr}-CQDs in water in Fig.~\ref{fig2}\textbf{d}) indicating more complex photodynamics in these cases.

The MPL contrast increases monotonically with the amplitude of the magnetic field, shown in Fig.~\ref{fig2}\textbf{e} for \textit{Gly}-CQDs dry and in water, and approaches saturation at the maximum field available in the experiment (46~mT). This trend is relatively consistent across samples, with a $B_{1/2}$ field (the field giving half of the saturation contrast) of approximately 20~mT (see full dataset in Fig. S15). 

To ensure the MPL contrast we observed is not due to an artefact such as mechanical coupling or heating, we repeated these measurements and looked at both the direct laser reflected off a clean glass coverslip as well as fluorescence from a calibration slide. Neither control measurement showed evidence of MPL (see SI Fig. S17). Additionally, we measured the amino acid starting material of all 19 samples and found most had no significant PL nor a magnetic field dependence. However, we tested other CQD samples which were found to exhibit MPL, see example of several commercially sourced CQDs in Fig. S17. This indicates that the observed MPL effect is not unique to CQDs derived from amino acids, although this synthesis technique presents important advantages in terms of simplicity and scalability. 

\begin{figure}
    \centering
    \includegraphics[width=\linewidth]{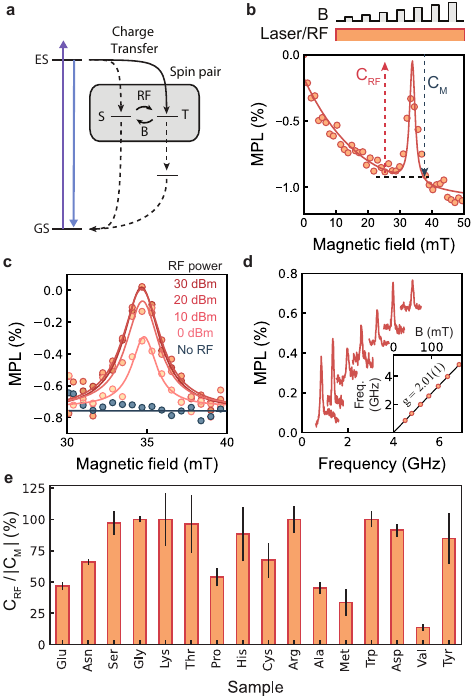}
    \caption{\textbf{Optically detected magnetic resonance and spin-pair origin.}
    \textbf{a} Simplified model with optical cycling between ground and excited states (GS and ES). Charge transfer from the excited state creates a weakly coupled spin pair. The singlet state (S) of the spin pair decays directly back to the GS, while the triplet state (T) decays to an intermediate long-lived state first. Singlet-triplet mixing occurs either at $B=0$ or under RF driving, favouring a rapid return to the GS, resulting in maximum PL. 
    \textbf{b} PL change relative to the $B=0$ condition, as a function of magnetic field with an RF driving field applied at 980~MHz, measured for dry \textit{Gly}-CQDs. A cancelling of the MPL effect is evident at $B\approx35$\,mT corresponding to the ESR condition. The solid line is a fit with a stretched-exponential decay plus a Lorentzian peak.
    \textbf{c} Zoom-in on the ESR peak for different RF driving powers. The solid line is a Lorentzian fit. 
    \textbf{d} ESR spectra taken by scanning the RF frequency at different magnetic field strengths from 30 to 170\,mT. 
    Inset: Relationship between the ESR position and the applied magnetic field, yielding a slope $g=2.01(1)$. 
    \textbf{e} Contrast of the ESR peak ($C_{\rm RF}$) as a fraction of the MPL contrast at the same field ($C_{\rm M}$), i.e. the ratio $C_{\rm RF}/|C_{\rm M}|$, for the 16 \textit{aa}-CQDs samples exhibiting clear MPL contrast, measured at $B=35$\,mT. For most samples the ratio approaches 100\% meaning the MPL effect is completely suppressed by the RF. 
    }
    \label{fig3}
\end{figure}

The observed MPL effect is indicative of spin-dependent photodynamics. We postulate that the underlying mechanism is similar to the so-called radical pair mechanism, whereby charge transfer within a CQD particle following photoexcitation creates a metastable weakly coupled spin pair ~\cite{evans_magnetic_2013,steiner_magnetic_1989, boehme_theory_2003, mclauchlan_invited_1991} (Fig.~\ref{fig3}\textbf{a}). The singlet and triplet states of the spin pair recombine to the ground state at different rates due to spin selection rules. Without a magnetic field, the singlet and triplet are spin mixed by hyperfine interactions, resulting in rapid recombination and the maximum PL emission rate. When a magnetic field is applied with sufficient strength (exceeding these hyperfine interactions) to lift the degeneracy, upon optical cycling the triplet state becomes preferentially populated and decays to a long-lived intermediate state (lifetime of several seconds in our case), leading to a drop in PL rate. This mechanism is common across a variety of molecular and solid-state systems~\cite{evans_magnetic_2013, xiang_mechanism_2025,shinar_optically_2012, robertson_universal_2024}, and explains the observed multi-component dynamics and the range of $B_{1/2}$ fields (corresponding to the strength of hyperfine interactions). 

To demonstrate that this is the origin of the MPL response, we directly drive the spin system using an RF field of frequency $f$. For a weakly coupled spin pair, an electron spin resonance (ESR) is expected when $hf=g\mu_{\text{B}}B$ where $h$ is Planck's constant, $g$ the Land{\'e} $g$-factor, $\mu_{\text{B}}$ the Bohr magneton, and $B$ the magnetic field.  
Using dry \textit{Gly}-CQDs as a test sample, we begin by applying a constant RF frequency of $f = 980$~MHz while simultaneously ramping the applied magnetic field (Fig.~\ref{fig3}\textbf{b}). 
We observe a peak in the MPL spectrum at the expected ESR condition of $B=hf/g\mu_B\approx35$\,mT. This ESR peak is explained by the resonant RF field mixing the singlet and triplet states, similar to the $B=0$ condition, thus cancelling the MPL effect. 
This is direct evidence that the MPL effect comes from a demixing of spin states, which can be reversed by directly driving the spins themselves. 
As shown in Fig.~\ref{fig3}\textbf{c}, the efficiency of this reversal depends on the driving RF power, which saturates (i.e. the PL is returned to the $B=0$ condition) at high power.
We can also scan the RF frequency, $f$, at a fixed magnetic field $B$ to see the ESR resonance, allowing us to apply larger fields using a permanent magnet. This is shown in Fig.~\ref{fig3}\textbf{d} where $B$ is increased up to 170 mT, which is accompanied by a shift in the resonance frequency  confirming the linear behaviour with $g=2.01(1)$ (see field calibration procedure in the SI Fig. S20) consistent with the weakly coupled spin pair model.
We were able to observe ESR for all 16 dry samples that exhibited MPL (see SI Fig. S19). At resonance, we find near complete PL recovery in most samples (Fig.~\ref{fig3}\textbf{e}), i.e. the amplitude of the ESR peak ($C_{\rm RF}$) matches the MPL contrast ($C_{\rm M}$) at the corresponding field. We suspect that the few samples that have incomplete recovery would require stronger RF driving fields. 
Additionally, by applying RF pulses we were able to observe coherent population oscillations of the spin pair (between singlet and triplet), indicating that it could be possible to apply more complex spin control sequences e.g. to measure the spin coherence time (see SI Fig. S21). 


\begin{figure}
    \centering
    \includegraphics[width=\linewidth]{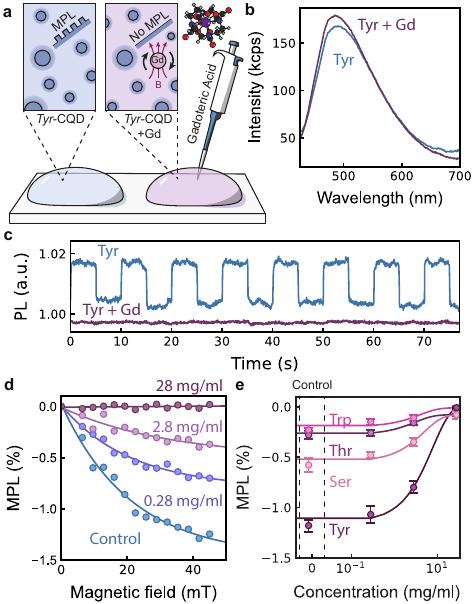}
    \caption{\textbf{Sensing paramagnetic molecules in solution.} 
    \textbf{a} Illustration of the paramagnetic sensing experiment, where gadoteric acid is added to a solution of CQDs causing a reduction in MPL contrast.  
    \textbf{b} PL emission spectrum with 405~nm excitation of the control \textit{Tyr}-CQD solution compared with 28 mg/ml of gadoteric acid, showing only a subtle difference. 
    \textbf{c} MPL trace when switching $B$ between 0 and 46\,mT every 5\,s, for the \textit{Tyr}-CQD control (blue) and with 28 mg/ml of gadoteric acid (purple).
    \textbf{d} MPL response (\textit{Tyr}-CQD) to increasing magnetic fields for different gadoteric acid concentrations (solids lines are exponential fits). 
    \textbf{e} MPL contrast at $B=46$\,mT of several \textit{aa}-CQDs as a function of gadoteric acid concentration (solids lines are exponential fits).
    }
    \label{fig:sense}
\end{figure}
Having established that a spin pair structure is responsible for the MPL observed in our samples, we next consider how this understanding may enable new sensing modalities.
CQDs are generally known to be very sensitive to the local electric charge environment which can alters the spectral emission significantly \cite{jafarmolaei_principles_2020}.
Our \textit{aa}-CQDs are no exception, with significant spectral and PL intensity changes to pH and different ionic environments (see Fig.\,S22).
However, the fact that we see spin-dependent PL changes means that these CQDs should also be sensitive to locally fluctuating magnetic fields e.g. from paramagnetic molecules \cite{Steinert2013, robertson_detection_2023, Ermakova2013, Kaufmann2013}, even if those do not electrically interact with the CQDs (Fig.~\ref{fig:sense}\textbf{a}). As an example, we consider the effect of chelated Gd ions in gadoteric acid, a common spin contrast agent, on the four best performing CQD samples. 

Interestingly, the effect on the spectrum of \textit{Tyr}-CQDs (Fig.~\ref{fig:sense}\textbf{b}) is minimal, making it difficult to separate from experimental drifts. In contrast, the suppression of the MPL effect from Gd ions is significant, with any signature of MPL removed at 28 mg/ml (Fig.~\ref{fig:sense}\textbf{c}). By ramping the magnetic field (Fig.~\ref{fig:sense}\textbf{d}), we confirm that the MPL contrast gradually decreases with increasing Gd concentration while the $B_{1/2}$ field remains relatively unchanged ($\approx20$\,mT for these \textit{Tyr}-CQDs). 
An estimate of sensitivity to changes in Gd concentration is given by
\begin{equation}
    \eta = \frac{1}{S \sqrt{\alpha R}} 
\end{equation}
where $\mathcal{S}$ is the gradient of MPL signal versus Gd concentration, $R$ is the PL rate and $\alpha$ is the duty cycle of the measurement ($\alpha = 0.5$ in these measurements).
This gives a sensitivity of 10-30 $\mu$g/ml$\sqrt{\rm Hz}$ for all 4 CQD. 

The reduction in MPL contrast with gadoteric acid occurs because the Gd ions act as a broadband magnetic noise source~\cite{Steinert2013} which incoherently drives the CQD spins in a similar fashion to the RF driving shown in Fig.~\ref{fig3} but over the entire magnetic field range rather than in a specific region. 
Moreover, the suppression of MPL contrast (at $B=46$\,mT) is detected in all of the four \textit{aa}-CQDs tested (Fig.~\ref{fig:sense}\textbf{d}), confirming the robustness of the effect. Importantly, the MPL effect, when measured as an absolute PL change, is insensitive to variations in background PL. This makes MPL a potentially robust indicator of the presence of paramagnetic molecules in high-background scenarios, e.g. in biosensing applications where cell autofluorescence is often problematic. In Fig. S23, we present a demonstration of this idea where the MPL contrast is extracted despite large background variations.

\section*{\label{Conclusion}Conclusion}

In this work, we have shown that amino-acid derived CQDs commonly host optically active electronic spin states amenable to manipulation with magnetic and RF fields. We observed magnetic modulation of the PL from 16 of the 19 samples studied, with magnetically induced contrasts in the range of $\approx$-0.1 to -1.5\% from dry samples. This effect persisted in suspension.

We showed that the MPL effect is indeed a result of unpaired electronic spins by direct manipulation using resonant RF fields to study the ESR behaviour as a function of magnetic field strength. This approach was used to measure a $g$-factor of $g\approx2$ for a representative sample. In combination with the fact that the ESR contrast is of opposite sign and matching amplitude to the MPL contrast, a radical pair mechanism is likely at the origin of the spin-dependent PL. Finally, we demonstrated the possible utility of these \textit{aa}-CQDs for sensing by measuring changes to the MPL contrast in the presence of a paramagnetic species.

This work shows the potential of CQDs for applications beyond their current uses, and offers an intriguing alternative to established spin-based sensors such as diamond nanoparticles. The key advantage of CQDs is the scalability and ease of synthesis, as well as their biocompatibility. However, the MPL contrast is comparatively low and will need to be improved to make CQDs competitive in terms of sensitivity. Recent progress in other optical-spin systems suggests potential avenues to increase the MPL contrast via materials engineering or multi-colour or pulsed excitation~\cite{mena_room-temperature_2024,wang_giant_2024,feder_protein_2025}.

With improvement, we foresee the MPL behaviour of CQDs being used as an additional readout modality in intracellular experiments. For example, as a sensor of paramagnetic species like bio-available iron~\cite{grantNonmonotonicSuperparamagneticBehavior2023}, or as a method for background subtracted imaging using lock-in methods (see SI Fig. S23 for a simple demonstration). Crucially, these measurement schemes are relatively straightforward to implement and can work alongside existing sensing modalities relying on PL intensity and spectral shifts.

\section*{Experimental Section}

\textit{Synthesis of CQDs:}
CQDs were synthesized via pyrolysis of an amino acid precursor. Briefly, 2~g of the selected amino acid was placed into a 100~mL round-bottom Pyrex glass flask and heated using a magnetic hotplate stirrer. The precursor was pyrolyzed within a temperature range of 210–340°C, depending on the melting point of the amino acid, for 2 hours without stirring. Heating was discontinued, allowing the reaction mixture to cool naturally to approximately 150°C. Subsequently, 30~mL of Milli-Q water was added, and the mixture was vigorously stirred at 1000~rpm while cooling further to room temperature. The resulting suspension was transferred to 50~mL plastic centrifuge tubes and centrifuged at 6000~rpm for 10~min. The supernatant containing dispersed CQDs was carefully collected and sequentially filtered twice through a 0.22~$\mu$m cellulose syringe filter, yielding a homogeneous filtrate. The filtrate was then transferred into a 30~mL glass vial and frozen at -20°C for 24~h. Following thawing at room temperature, any precipitated agglomerates were removed through decantation and additional filtration. This freeze-thaw purification cycle was typically repeated twice or until no visible deposits remained upon thawing. Finally, the purified CQD solution was freeze-dried at -90°C for 24~h, resulting in dry CQD powder used for subsequent characterization.

X-ray photoelectron spectroscopy was performed on a Thermo Scientific K-alpha XPS equipped with Al K-$\alpha$ source (1486.7 eV). Spectra were collected using an X-ray power of 72W and a pass energy of 50 eV. A low energy electron/Ar$^+$ ion source was used for charge neutralisation. Data analysis was perfomed using CasaXPS software v2.3.
TEM imaging was performed on a JEOL F200 CEFG operating at 200 kV. A Gatan Rio16 camera was used for image capture. UV-vis absorbance spectra were collected using an Agilent Cary 5000 UV–vis–NIR spectrophotometer.
For XRD data collection, a Rigaku SmartLab, equipped with a rotating anode CuK$\alpha$ source (45kV, 200mA), and Hypix detector, was employed to obtain grazing incidence data from dried CQD samples. 
The CQD solutions had been previously dropcasted on a clean glass substrated and air-dried at ambient. 
Data was collected from 5$^\circ$ to 60$^\circ$ 2$\theta$ with a step size of 0.04 deg, $\omega$ set at a glancing angle of 0.183$^\circ$, at a scan rate of 0.5$^\circ$ per minute. 
The incidence slit, beam limiting mask and $\omega$ setting combined to maintain a beam footprint of $\sim$12mm by $\sim$15mm. 
Analyses were performed on the collected XRD data using the Bruker XRD search match program EVA™ 7.

\textit{MPL and ESR measurements:}
For the MPL and ESR measurements, the CQD powder was dissolved in MilliQ water to a concentration between 1-3~mg/mL. Dry samples were then created by dispensing 7\,$\mu$L onto a clean glass coverslip and leaving to dry under ambient conditions inside a fume hood. Suspensions were prepared by dispensing 7\,$\mu$L into a SecureSeal Imaging Spacer (SS1X9) held between two coverslips. For the sensing measurements presented in Fig. 4, serial 1 in 10 dilutions of gadoteric acid (starting concentration 280 mg/mL) were made using each sample of \textit{aa}-CQD as the diluent. Final samples were made by dispensing 7\,$\mu$L of each mixture into a spacer/coverslip assembly as above.

The MPL and ESR measurements were performed using a custom-made widefield microscope system as shown in Figure~S10 of the SI.
The optical excitation was from a 405~nm CW laser (Hubner Photonics Cobolt 06-01 Series). The laser beam was focused onto the back aperture of a 20x objective (Nikon S Plan Fluor ELWD 20x, NA = 0.45) using a 200~mm lens, forming a $100\,\mu$m diameter spot on the sample. The laser power used in the measurements was 10\,mW typically, i.e. an intensity of order 100\,W/cm$^2$. 
The excitation and collection were separated using a dichroic mirror (Semrock Di02-R405-25x36) and further filtered using a longpass filter with a 420\,nm cuton wavelength (semrock LP02-405RU-25) and imaged using a sCMOS camera (Andor Zyla 5.5-W USB3).

The magnetic field was applied using an electromagnet (APW EM100-12-222-A) connected to a programmable power supply (Keithley SMU2450), with a voltage of 20 V (450 mA current) corresponding to a field of 46 mT at the sample as calibrated using a reference hBN sample \cite{scholten_multi-species_2024}. The power supply was used in constant voltage mode for the MPL switching measurements (switching between 0 and 20 V), or in constant current mode for the $B$ ramps (including for ESR spectroscopy) to ensure field stability.  

RF excitation was provided by a signal generator
(Windfreak SynthNV PRO), amplified (Mini-Circuits HPA-50W-63+), and fed into a PCB equipped with a coplanar waveguide. The coverslip with the CQD sample was placed directly on top of the PCB. 

A custom Labview program was used to acquire camera images while controlling the power supply and RF signal generator. The pixels exposed by the laser were summed together to form the PL signal displayed throughout the paper. The raw PL traces were processed using custom scripts to remove the background variations (see Fig.\,S11 of the SI).

All measurements were performed at room temperature under an ambient atmosphere.\\

\noindent\textbf{Acknowledgements}\\
We thank Rick Franich for providing the gadoteric acid, and Sherman Wong (CSIRO) for assistance with XRD analysis. This work was supported by the Australian Research Council (ARC) through grants DE230100192 and DP250100973. The work was performed in part at the RMIT Micro Nano Research Facility (MNRF) in the Victorian Node of the Australian National Fabrication Facility (ANFF) and the RMIT Microscopy and Microanalysis Facility (RMMF). I.O.R. is supported by an Australian Government Research Training Program Scholarship. P.R. acknowledges support through an RMIT University Vice-Chancellor’s Research Fellowship. \\

\noindent\textbf{Conflict of Interest}\\
The authors declare no conflict of interest.\\

\noindent\textbf{Data Availability Statement}\\
The data that support the findings of this study are available from the corresponding author upon reasonable request.\\

\noindent\textbf{Keywords}\\
Carbon quantum dots, quantum sensing, quantum materials, nanomaterials

\bibliographystyle{MSP}
\bibliography{bib.bib}

\clearpage

\onecolumngrid
%

\section{Additional structural characterisation data}

\begin{figure*}[b]
	\centering
	\includegraphics[width=0.7\linewidth]{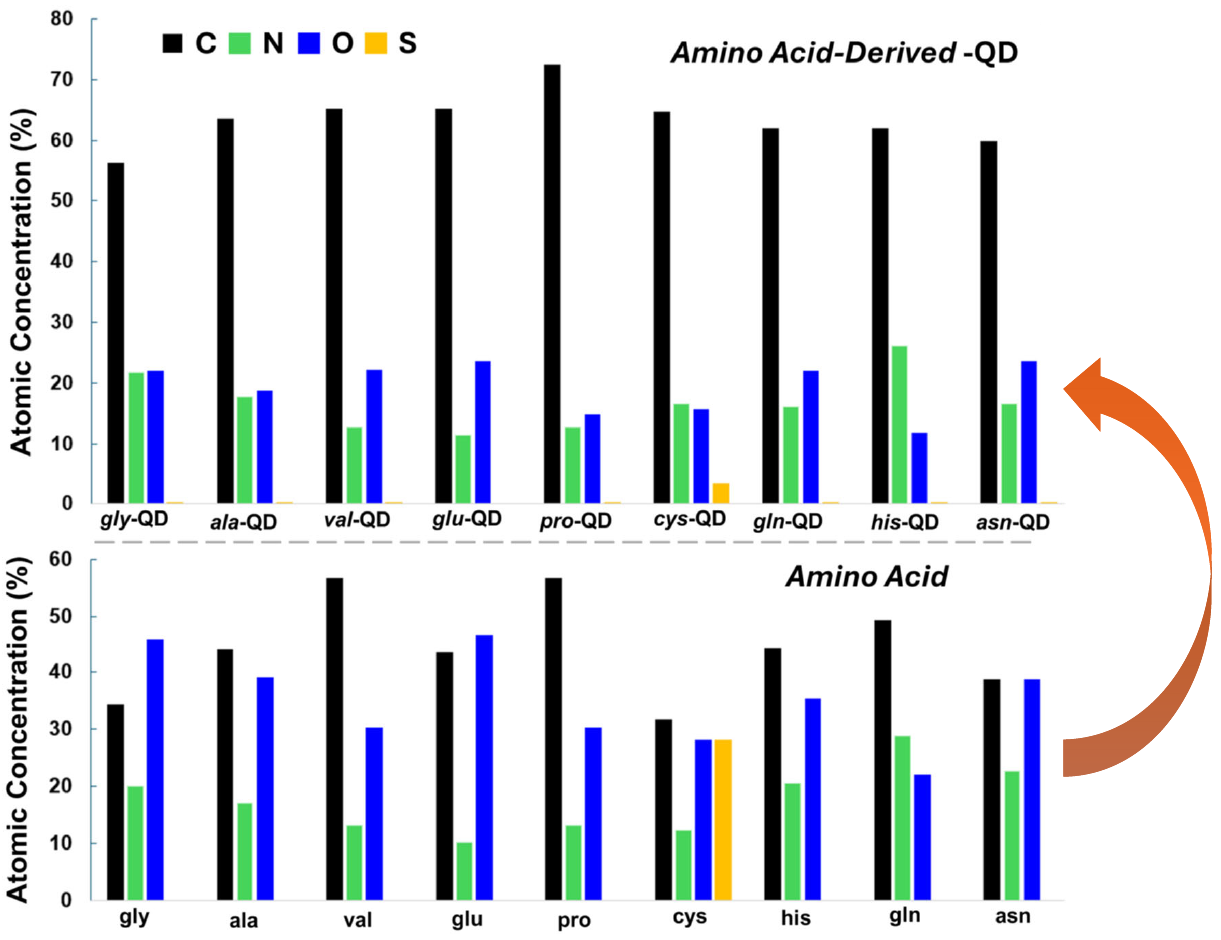}
	\caption{Bar graphs showing the atomic concentrations of elements based on XPS analysis of \textit{aa}-CQDs (top) compared to the elemental compositions of amino acid precursors (bottom).}
	\label{SI_XPS}
\end{figure*}

Elemental analysis by X-ray photoelectron spectroscopy (XPS) before and after pyrolysis is shown in Figure~\ref{SI_XPS} for 9 of the amino acids studied. It reveals compositional changes during the formation of the \textit{aa}-CQDs, with increased atomic C and reduced O, N, and S content compared to their amino acid precursors, indicating dehydration and volatilisation of heteroatom-containing species. Only the \textit{asn}-CQD sample deviates from this trend, showing reduced atomic C and increased O concentrations, suggesting predominant release of carbon oxides. The release of small molecular species during amino acid pyrolysis drives structural rearrangements such as cyclisation and increased unsaturation, ultimately yielding a graphitic carbon core (as evidenced by the XRD of gly-CQD, main text) that is decorated with diverse functional groups.

\begin{figure*}
	\centering
	\includegraphics[width=\linewidth]{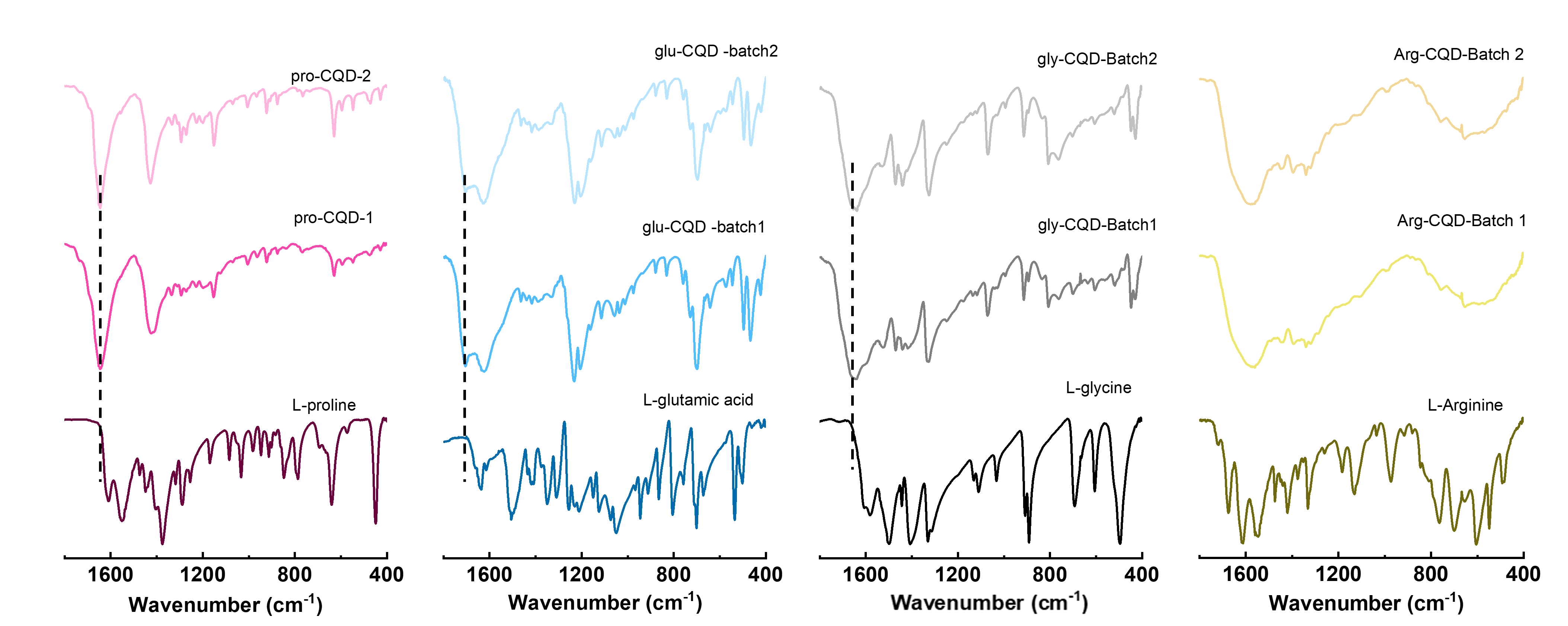}
	\caption{FITR spectra of the animo acid precursor and two separate synthesis of carbon quantum dots for Proline, Glutamatic acid, Glycine and Arginine. }
	\label{SI_FTIR}
\end{figure*}

The Fourier-transform infrared (FTIR) spectra of several precursors and their CQD counterpart are presented in Fig.~\ref{SI_FTIR}, as an illustrative example highlighting the chemical transformation from the precursor to the CQD. 
This transformation is evidenced by the emergence of new peaks in the \textit{aa}-CQD spectrum, which are absent in the precursor spectrum and indicate the formation of new bonds and structural rearrangements. In the example of glycine, these spectra include the modification of both the sp$^2$ and sp$^3$ hybridization groups between the glycine and the \textit{gly}-CQD, as well as a notable absence in the \textit{gly}-CQD spectrum of the N-H bending vibration at 1500 cm$^{-1}$, observed in the glycine spectrum, confirming the amine groups' chemical transformation into new bonding environments. 
Additionally, multiple batches of several \textit{aa}-CQD synthesized indicate that their fingerprint FTIR spectra were similar (Fig.~\ref{SI_FTIR}). The pyrolysis appears to have yielded different materials due to the nature of the amino acid precursors. 


\begin{figure*}
	\centering
	\includegraphics[width=0.7\linewidth]{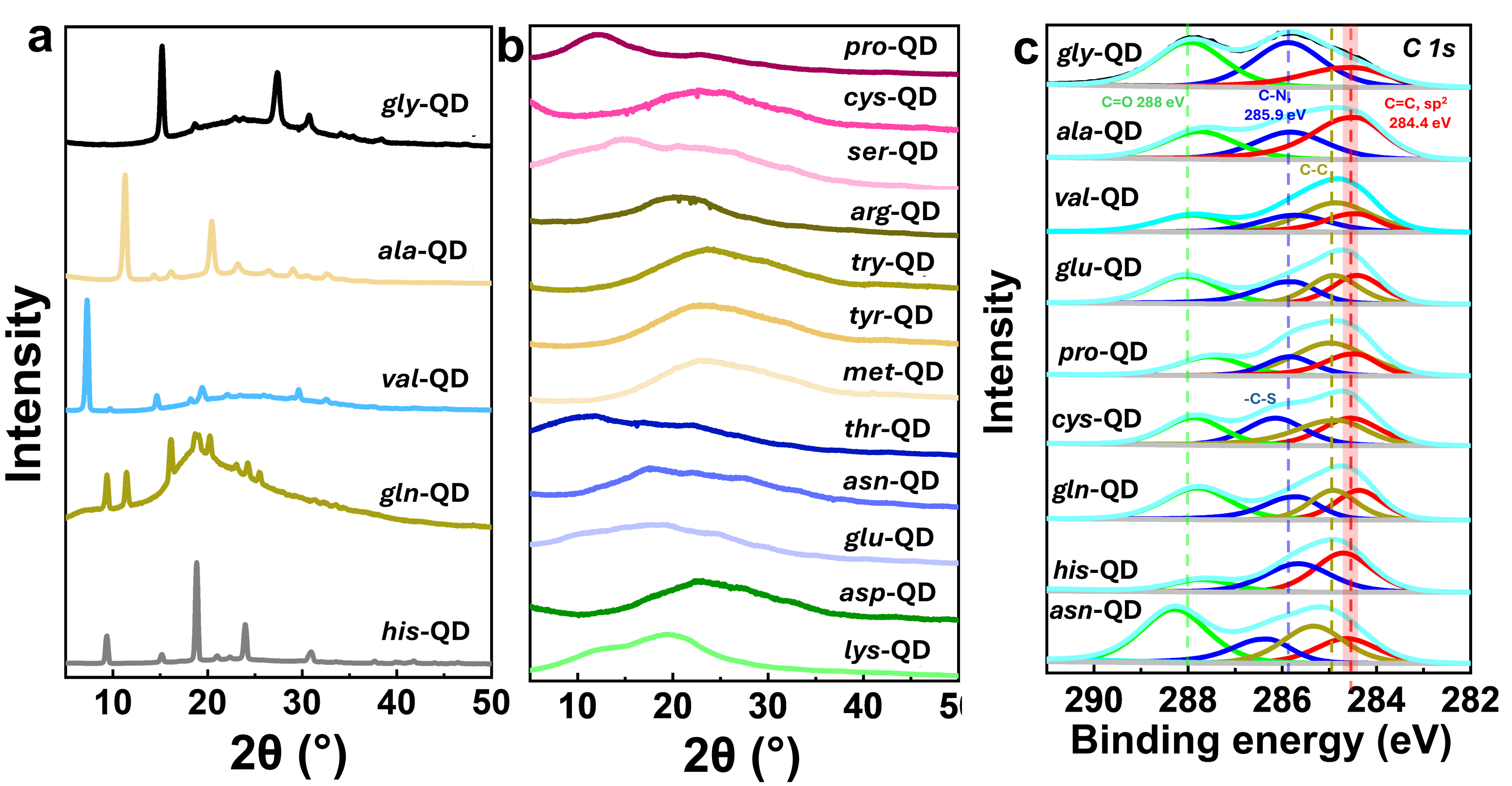}
	\caption{(a,b) Powder XRD patterns of (a) 5 \textit{aa}-CQD samples showing crystalline features, and (b) 12 other \textit{aa}-CQD samples showing an amorphous structure. (c) High-resolution C 1s XPS scan of 9 different \textit{aa}-CQD samples.}
	\label{SI_XRD}
\end{figure*}

Powder X-ray Diffraction (PXRD) was employed to determine the crystallinity of the carbon core in \textit{aa}-QDs. 
As shown in Fig. \ref{SI_XRD}a,b, only 5 CQD samples (derived from gly, ala, val, gln, and his) exhibit discernible crystalline features in their diffraction patterns, whereas the remaining samples display broad amorphous behaviour, which contrasts with the crystalline property of the precursors. 
The PXRD patterns shown in Fig. \ref{SI_XRD}a reveal sharper and intense diffraction peaks, which are uncommon for typical carbon quantum dots, generally characterized by broad, amorphous patterns due to their small diameters ($\approx$10~nm). 
In contrast, the observed diffraction peaks suggest the presence of larger or agglomerated crystalline structures within the carbon core. This behaviour aligns more closely with graphene and graphitic quantum dots, which typically have larger sizes ($\approx$20 nm) and display sharp XRD peaks\cite{malode_4_2022}.
As an example, the diffraction pattern of \textit{gly}-CQD indicates the presence of two distinct phases, with prominent peaks at $2\theta = 15^\circ$ and $2\theta = 27.5^\circ$, corresponding to graphitic carbon nitride (g-C$_3$N$_4$) and graphene-like carbon, respectively. The formation of the g-C$_3$N$_4$ phase under relatively mild synthesis conditions is remarkable since it is typically associated with higher temperatures exceeding 500$^\circ$C \cite{ong_graphitic_2016}.

The high-resolution C 1s XPS spectra presented in Fig.~\ref{SI_XRD}c reveal characteristic features consistent with surface functionalization by oxygen- and nitrogen-containing groups. A peak centred at approximately 288.0 eV corresponds to the O=C– signal, indicative of carbonyl or carboxyl functional groups. 
The main hydrocarbon peak, located at around 286 eV, displays considerable broadening, which upon fitting, is attributed to distinct contributions from sp³-hybridised –C–N– species (286 eV), aliphatic –C–C– (~284.8 eV), and sp²-hybridised –C=C– (284.5 eV), as marked by the red-dotted line in Fig.~\ref{SI_XRD}c. 
A notable shift in the -C-N- region is observed for \textit{cys}-CQD, with the corresponding peak shifting from 285.9 eV to 286.2 eV, suggesting the presence of C–S bonds. In the case of \textit{asn}-CQD, a marked shift of the C=O peak to higher binding energies is evident, consistent with protonation of carboxylate groups and reduced dehydration of the QD.


\begin{figure*}
	\centering
	\includegraphics[width=\linewidth]{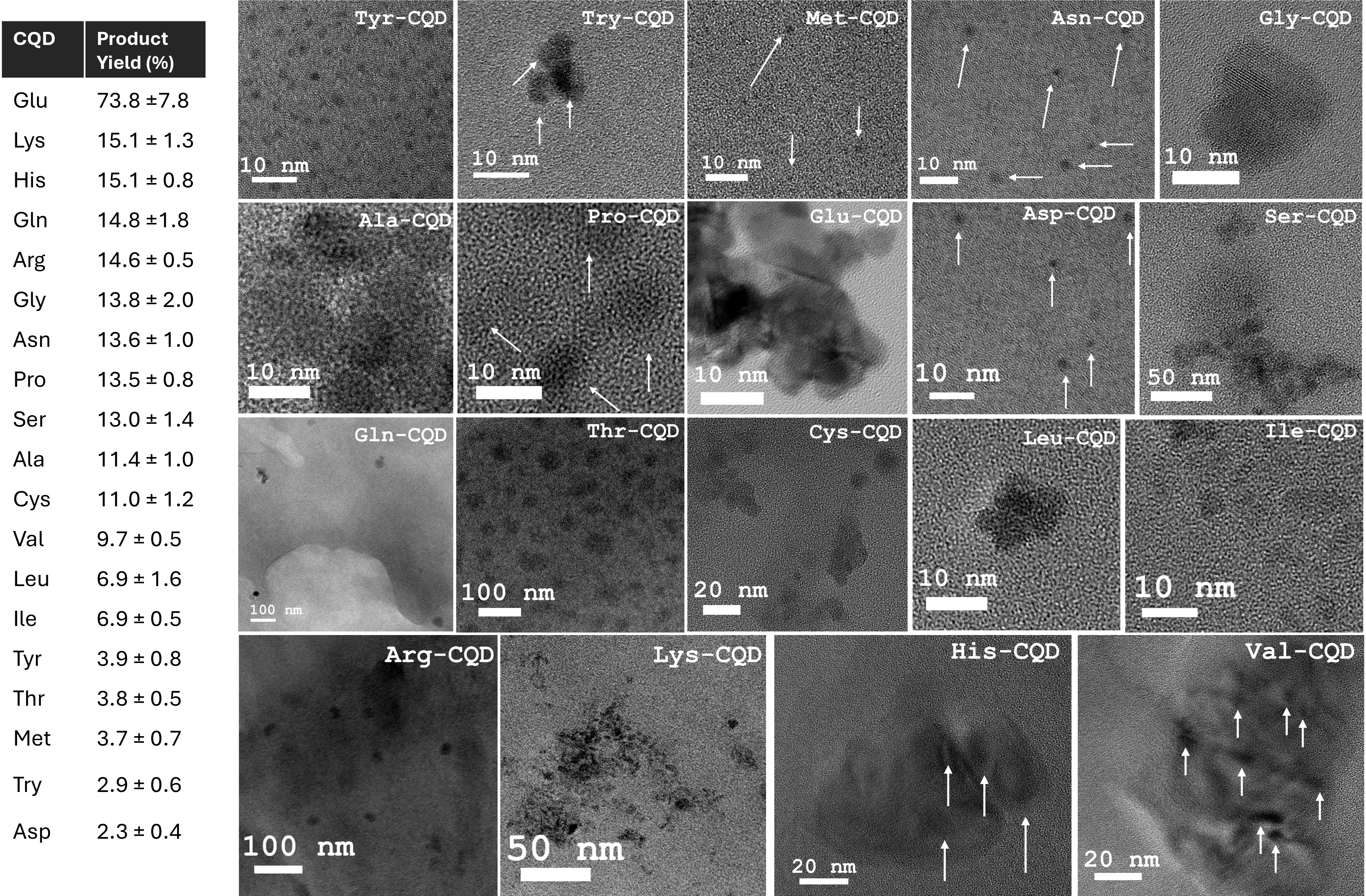}
	\caption{Transmission electron microscopy (TEM) images of the \textit{aa}-CQD samples with the estimated production yield.}
	\label{SI_TEM}
\end{figure*}

The transmission electron microscopy (TEM) images in Fig.~\ref{SI_TEM} complement the PXRD and XPS analyses by providing further insights into the morphology and crystallinity of the carbon cores in selected \textit{aa}-CQDs. 
TEM analysis reveals structural variations among the samples. The \textit{gly}-CQD sample featured agglomerated particles of approximately 15 nm with clearly defined lattice fringes and an interplanar spacing of approximately 0.38 nm, indicative of well-developed crystalline domains. In contrast, \textit{ala}-CQDs and \textit{pro}-CQDs display more uniform, spherical particles with an average size around 10 nm, suggesting moderately ordered structures. 
The \textit{glu}-CQDs sample, however, appears as smaller clusters of approximately 5 nm, reflecting a lower degree of crystallinity or the presence of highly disordered or fragmented graphitic structures. 
These morphological observations align with PXRD results, where sharper and more intense diffraction features \textit{gly}-CQD confirm enhanced crystallinity, while the broader, less-defined patterns observed for \textit{glu}-CQD suggest an amorphous or poorly ordered carbon phase.

We observed that CQDs derived from AA with hydrophilic side chains are preferentially extracted into Milli-Q water, yielding more than 12\%, with glu-CQDs achieving over 70\% after freeze-drying. In contrast, other CQDs, such as tyr-CQD, try-CQD, asp-CQD, and met-CQD that contain hydrophobic or aromatic side chains with at least two methylene groups, have less preference for water and remain at low concentrations, resulting in lower yields, as summarised in Fig.~\ref{SI_TEM}.

\section{Additional optical characterisation data}

\begin{figure*}
	\centering
	\includegraphics[width=0.6\linewidth]{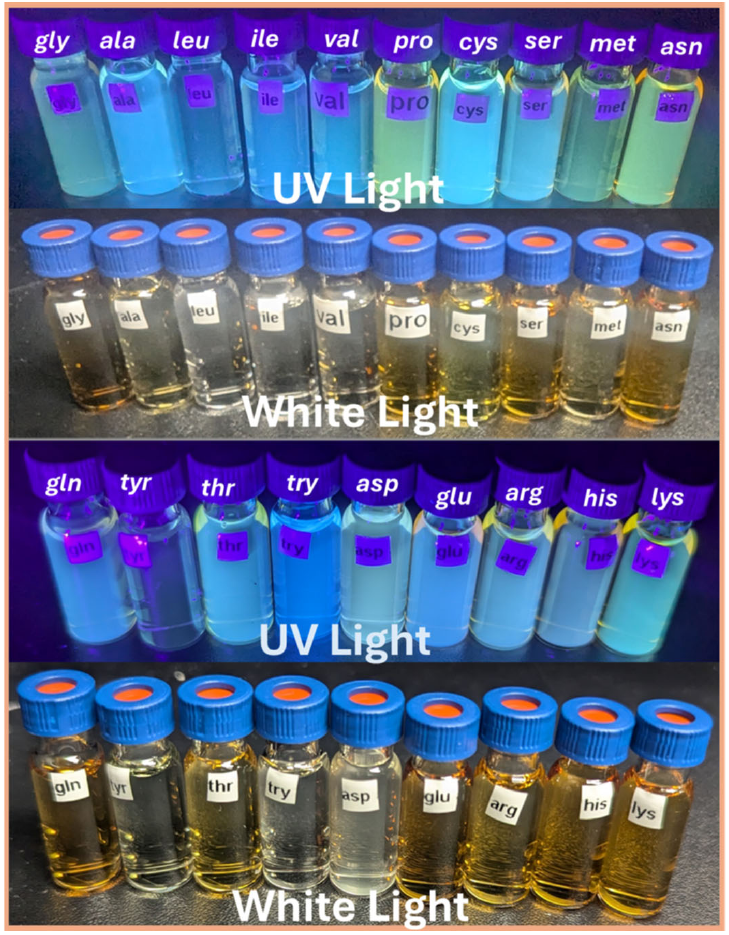}
	\caption{Photographs of \textit{aa}-CQD solutions under daylight and UV light irradiation.}
	\label{SI_CQDpics}
\end{figure*}

The dried \textit{aa}-CQDs exhibit excellent water solubility, which appear brownish yellow under white light but emit strong fluorescence, spanning from blue to green and greenish yellow under 320 nm UV-light irradiation, as illustrated in Fig. \ref{SI_CQDpics}.
Among the 20 amino acids subjected to pyrolytic treatment for the synthesis of \textit{aa}-CQDs, phenylalanine was the only precursor that failed to yield a viable product, resulting in an oily, water-immiscible residue; all other 19 samples were included in this study. 

\begin{figure*}
	\centering
	\includegraphics[width=0.8\linewidth]{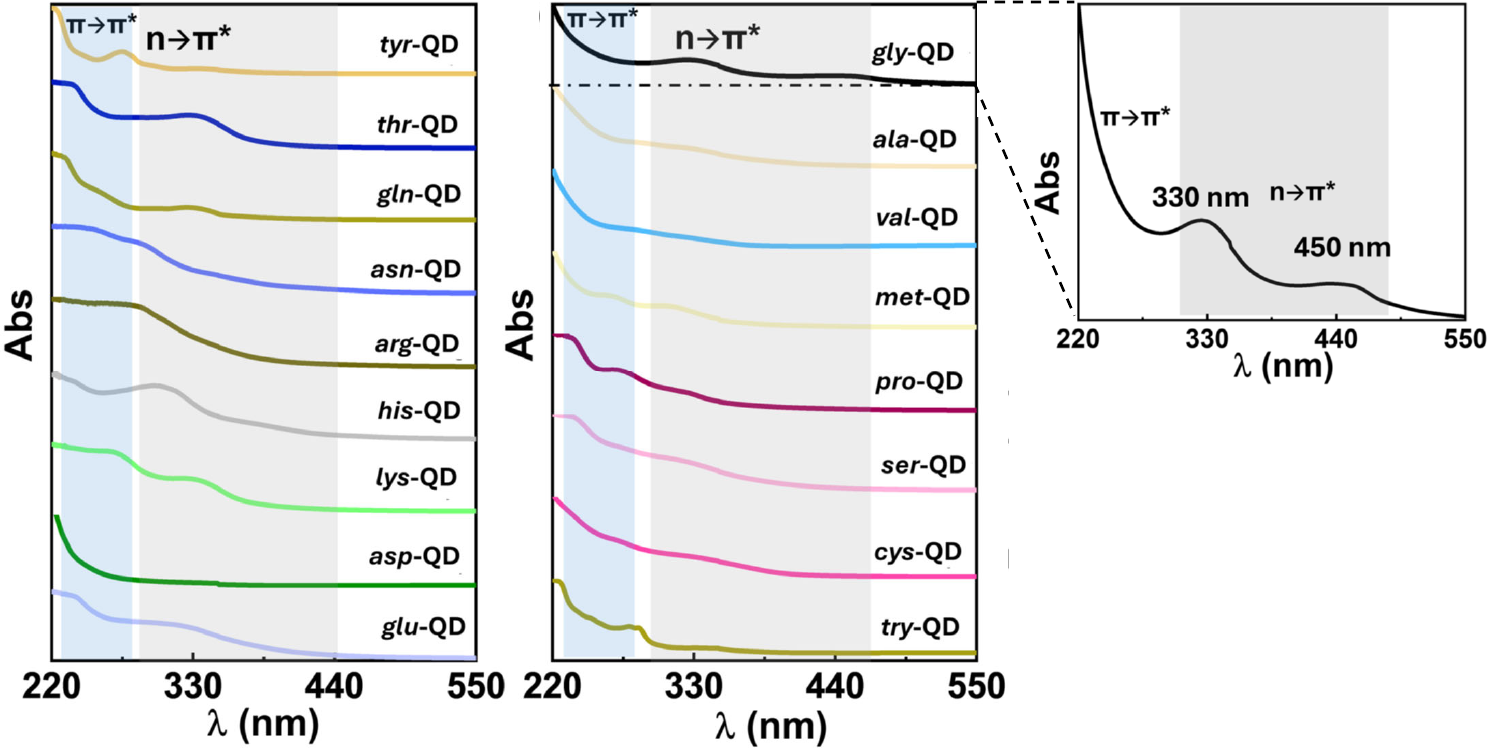}
	\caption{UV-vis absorbance spectra of \textit{aa}-CQDs derived from amino acids, indicating different regions of electronic transitions. Double peaks are observed in most samples except for \textit{asp}-CQD. The spectrum of \textit{gly}-CQD is shown in expanded view to visualize two absorption peaks at 330 nm and 450 nm.}
	\label{SI_absorption}
\end{figure*}

The UV-vis absorption spectra of the 19 \textit{aa}-CQDs presented in Fig. \ref{SI_absorption} reveal absorption peaks below 300 nm, indicating electronic $\pi-\pi^*$ transitions associated with aromatic sp$^2$-bonded carbon, typical of graphitic structures \cite{saxena_investigation_2011, mintz_deep_2021, wang_thickness-dependent_2014}. 
The \textit{gly}-CQD spectrum (inset on the right) further features 330 nm absorption, corresponding to n-$\pi^*$ transitions associated with -C=O and -C=N- functionalities, with an additional absorption band at 453 nm that indicates an extended conjugated network of sp$^2$-bonded carbon involved in n-$\pi^*$ transitions \cite{tonga_designing_2019}.


\begin{figure}
	\centering
	\includegraphics[width=0.7\linewidth]{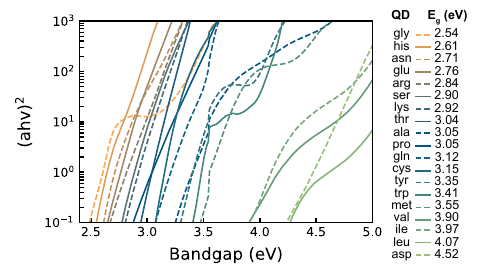}
	\caption{Tauc plots for estimating the size of the optically allowed direct bandgap of the \textit{aa}-CQDs. }
	\label{fig:tauc}
\end{figure}

The corresponding optical direct bandgap analysis, derived from the Tauc plot shown in Fig.~\ref{fig:tauc}, reveals variations in energy gaps ($E_g$) due to differences in the chemical composition and structures of the \textit{aa}-CQDs. Among these, \textit{gly}-CQD exhibits the lowest $E_g$ value of 2.54 eV, followed by \textit{his}-CQD, \textit{asn}-CQD, \textit{glu}-CQD, and \textit{arg}-CQD, which have bandgaps of 2.61 eV, 2.71 eV, 2.76 eV, and 2.84 eV, respectively. The influence of the side chains of amino acids, serving as sources of heteroatoms, coupled with the presence of unsaturation, is evident in their ability to tune the $E_g$ values of the resulting \textit{aa}-CQDs. This is significant since N-doping in carbon-based quantum dots has been demonstrated to shorten the bandgap. 
Furthermore, the presence of unsaturation, which preferentially facilitates extended conjugation for n-$\pi^*$ electronic transitions due to absorption edges at longer wavelengths, suggests why \textit{his}-CQD, derived from histidine, exhibits a shorter $E_g$ value of 2.61 eV compared to other \textit{aa}-CQDs, aside from \textit{gly}-CQD. Also, \textit{aa}-CQDs derived from amino acids with aliphatic side chains, such as \textit{val}-CQD, \textit{ile}-CQD and \textit{leu}-CQD seem to display the upper limits of the observed band gap range, with values of 3.90 eV, 3.97 eV, and 4.07 eV, respectively. This behaviour is attributed to the dominant sp$^3$ hybridization character inherent in their side chain structures.

\begin{figure}
	\centering
	\includegraphics[width=0.7\linewidth]{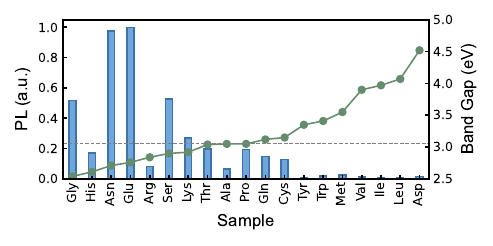}
	\caption{
		Comparison between PL intensity (left axis) and bandgap (right axis) for all \textit{aa}-CQD samples in water, ordered by increasing bandgap. The PL intensity was measured under the same conditions for all samples with 405\,nm laser excitation and a 420 nm longpass emission filter, and all samples in water had a similar concentration of CQDs. The horizontal dashed line indicates the bandgap corresponding to the laser line. }
	\label{SI_PLvsEg}
\end{figure}

In our MPL and ODMR experiments, we used a 405\,nm laser, corresponding to 3.06\,eV, which means the samples with a bandgap above this value will be less efficiently excited. The comparison between PL intensity and bandgap [Fig.~\ref{SI_PLvsEg}] show indeed that the 6 samples with the largest bandgaps (exceeding the laser energy) are significantly dimmer than the other samples.  
Broadly, we observe more PL from samples that have a lower optical bandgap with some notable exceptions which is likely related to differences in the size, structure, or surface chemistry of the respective CQDs. 
Interestingly, we are still able to detect the PL above our background noise even for the dimmest sample (\textit{Tyr}-CQD) which is 200 times weaker than the brighest (\textit{Glu}-CQD).
Therefore, the subsequent measurement will be performed on all 19 \textit{aa}-CQD samples. The results will be presented in order of decreasing brightness.  

\begin{figure}
	\centering
	\includegraphics{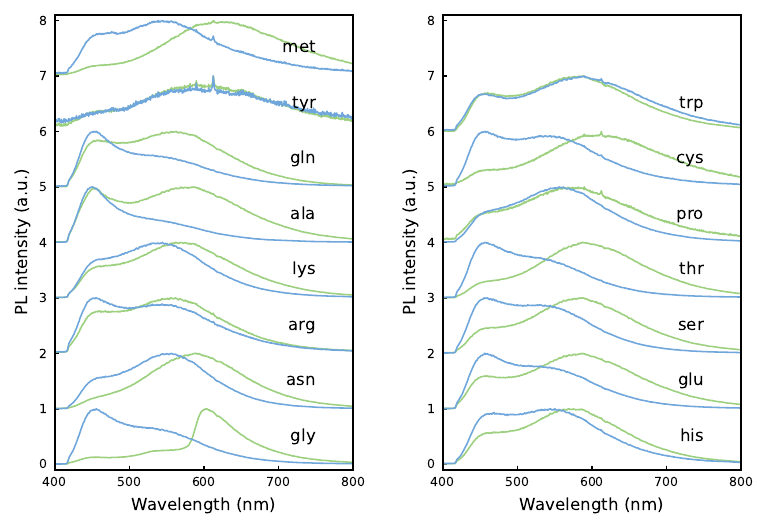}
	\caption{
		PL spectra of 15 of the \textit{aa}-CQD samples both dry (green) and in water (blue), under 405\,nm laser excitation with a 420\,nm longpass filter. The sharp peaks are artefacts from the spectrometer.}
	\label{SI_PLspectra}
\end{figure}

As expected from the large variations in structural properties, surface chemistry, and optical absorption spectra, we observe great variations in the PL emission spectra between the different \textit{aa}-CQD samples, taken under 405\,nm laser excitation [Fig.~\ref{SI_PLspectra}].
Generally, there is a red shift in the dry phase compared to in water, with most dry samples peaking at around 600\,nm compared to 450-500\,nm for the samples in water. Exceptions are trp and tyr which behave in a similar fashion in both phases. 
Interestingly, both of these CQD samples also have the best MPL contrast in water and are one of the few samples that perform better in water compared with dried (Fig.~\ref{SI_MPL vsB}).

\section{Additional magneto-PL data and methods}

\subsection{Experimental apparatus}

\begin{figure}
	\centering
	\includegraphics[width=0.9\linewidth]{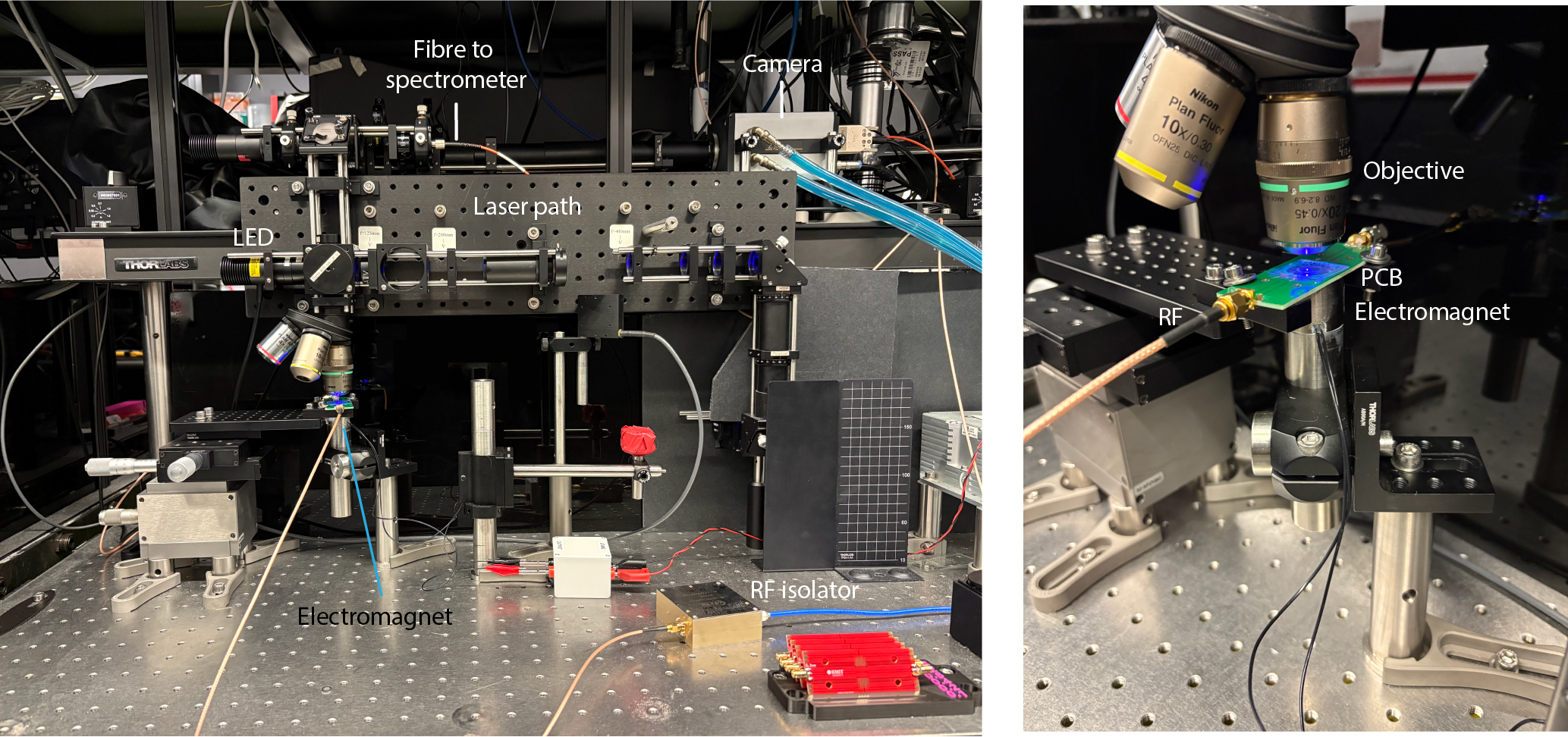}
	\caption{
		Images of the custom-built widefield fluorescence microscope system used for the MPL and ODMR measurements.}
	\label{fig:exp cam}
\end{figure}

The setup used for the MPL and ODMR measurements is described in the Method section of the main text. Photographs of the setup are shown in Fig. \ref{fig:exp cam}.

\subsection{Background subtraction}
\begin{figure}
	\centering
	\includegraphics[width=\linewidth]{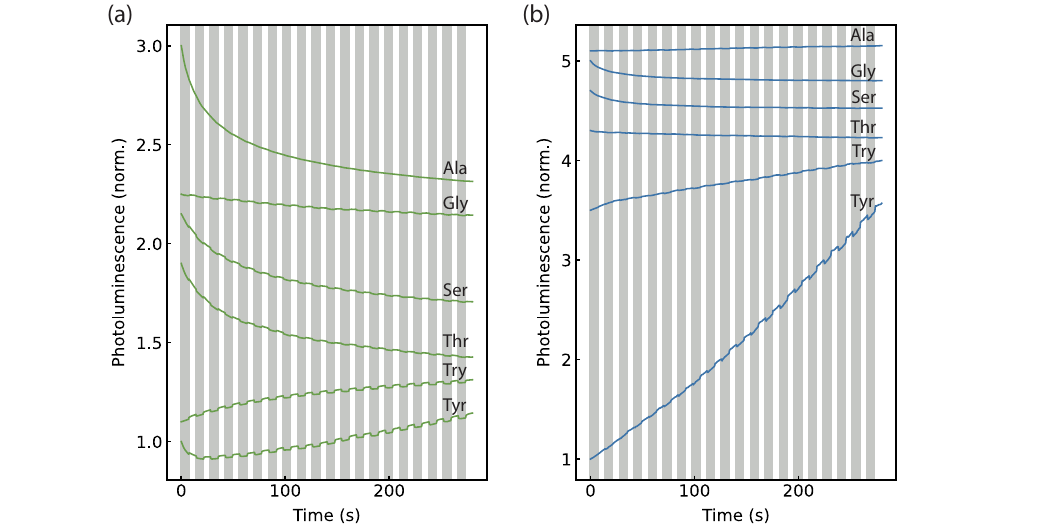}
	\caption{(a) Example raw MPL trace measured for the dry \textit{aa}-CQDs samples shown in the main text where the magnetic field ($B=46$\,mT) is switched on and off every 7\,s. (b) The same samples again, in water. These traces demonstrate the varied PL behaviour under continuous illumination.}
	\label{SI_bkg_raw}
\end{figure}

\begin{figure}
	\centering
	\includegraphics[width=\linewidth]{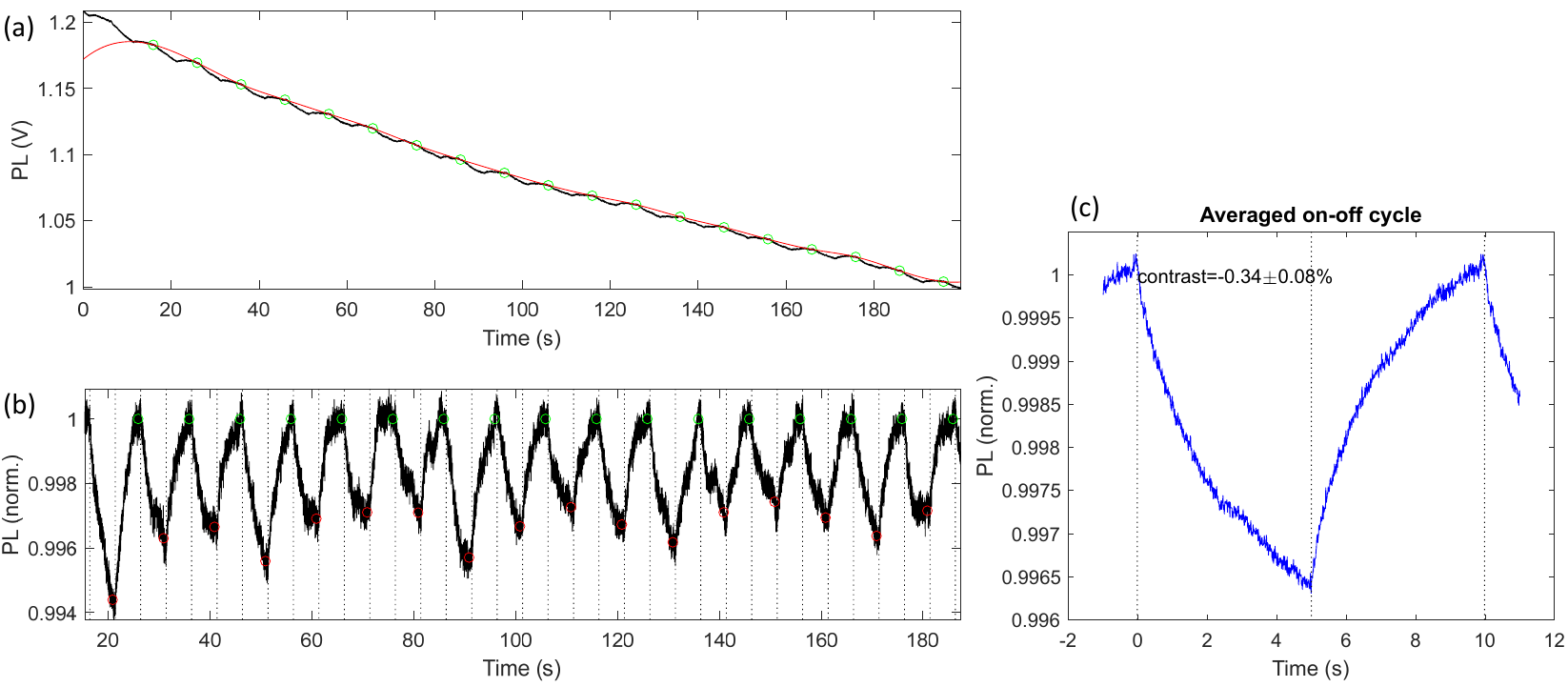}
	\caption{(a) Example raw MPL trace measured for dry \textit{his}-CQDs, where the magnetic field ($B=46$\,mT) is switched on and off every 5\,s. The green circles represent the regions near the end of each off period used to estimate the background variation (red line). (b) MPL trace after removing the background variation, i.e. ${\rm PL}_{\rm norm}={\rm PL}_{\rm raw}/{\rm PL}_{\rm bkg}$. The green (red) circles indicate the regions near the end of each off (on) period uased to estimate the MPL contrast. The vertical dotted lines represent the magnet switching times as determined by the control software. (c) Average on-off cycle obtained by averaging all the cycles measured in (b).}
	\label{SI_bkg}
\end{figure}

Typically, though not always, the \textit{aa}-CQDs were seen to photobleach under continuous laser illumination. To facilitate the observation of the MPL effect, the background PL variation ${\rm PL}_{\rm bkg}$ (i.e. independent of magnetic field effects) was removed from the raw ${\rm PL}_{\rm raw}$ to obtain a normalised trace ${\rm PL}_{\rm norm}={\rm PL}_{\rm raw}/{\rm PL}_{\rm bkg}$. ${\rm PL}_{\rm bkg}$ was estimated by taking the mean of the raw PL in the last 20\% of each magnet-off period and interpolating with a spline function, as illustrated in Fig. \ref{SI_bkg}a. 

In the normalised trace [Fig. \ref{SI_bkg}b], we can similarly take the mean of the last 20\% of the magnet-on periods and deduce the relative MPL contrast, $C_M={\rm PL}_{\rm on}/{\rm PL}_{\rm off} - 1$, which gives $C_M=-0.34\pm0.08\%$ in this example corresponding to the mean and standard deviation from all the on-off cycles measured. Note that when the PL has not fully saturated within the allowed time, as is the case in the example of Fig. \ref{SI_bkg}c, this definition underestimates the maximum MPL contrast that could be achieved with a longer settling time between switching events.    

\subsection{Temporal dynamics of MPL switching}

\begin{figure}
	\centering
	\includegraphics{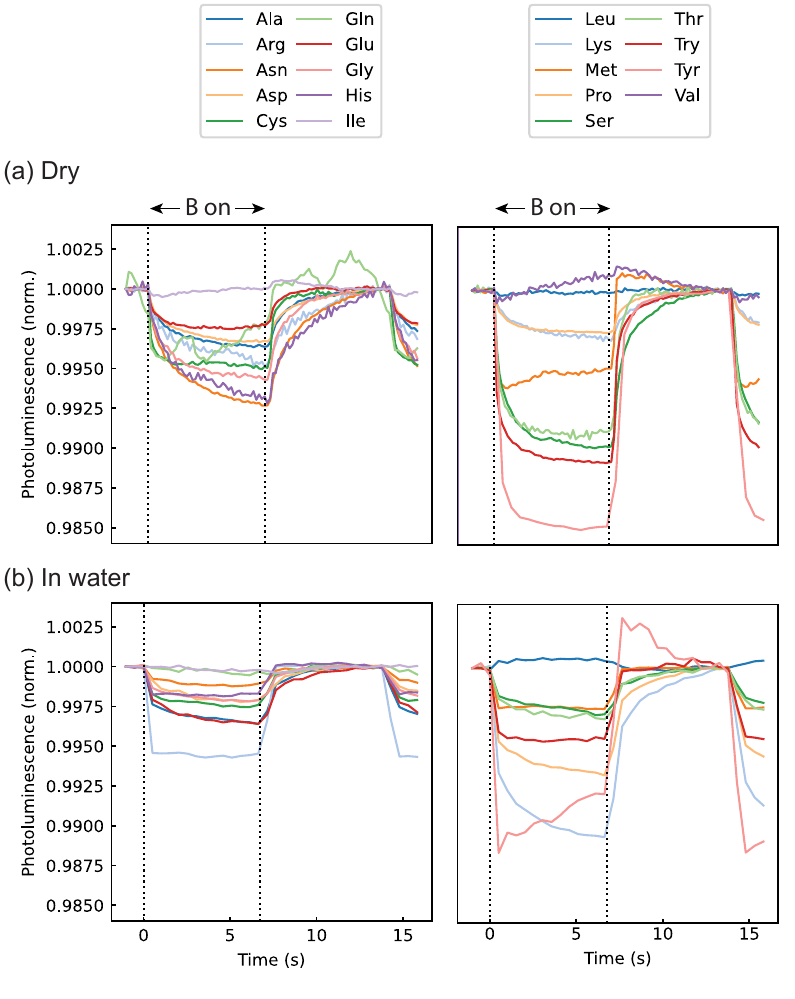}
	\caption{
		Decay curves of the PL as the magnetic field ($B=46$\,mT) is switched on, averaged from multiple on-off cycles as in Fig. \ref{SI_bkg}c, for (a) the dry samples and (b) in water.  
	}
	\label{SI_MPL_All}
\end{figure}

\begin{figure}
	\centering
	\includegraphics{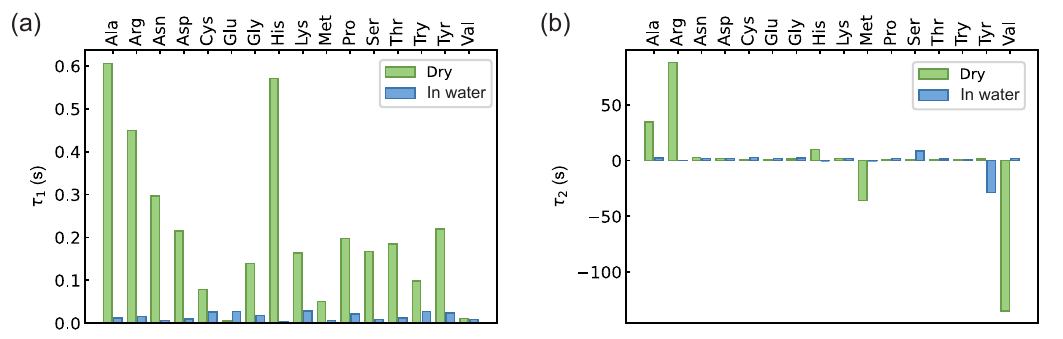}
	\caption{
		(a) The short decay time, $\tau_1$, scale from the MPL averages of dry (green) and in water (blue) data. (b) The long decay time, $\tau_2$ scale from the MPL averages of dry (green) and in water (blue) data.
	}
	\label{SI_Taus_aaCQDs}
\end{figure}

To visualise the time dynamics of the PL change as the magnetic field is switched on and off, we re-plot the average on-off cycle as in Fig. \ref{SI_bkg}(c). The resulting decay curves are shown in Fig. \ref{SI_MPL_All} for all of the \textit{aa}-CQD samples dry (a) and in water (b). It is clear that there are a broad range of dynamics across the samples. We fit these dynamics with a biexponential of the form:
\begin{equation}
	I(t) = C_1\exp(-t/\tau_1)+C_2\exp(-t/\tau_2)+I(\infty)
\end{equation}
where $\tau_1$ and $\tau_2$ characterise the short and long decay times respectively. The values of $\tau_1$ and $\tau_2$ are shown in Fig. \ref{SI_Taus_aaCQDs}(a) and (b) respectively. In almost all cases both the short and long decay times are longer in the dried samples, however the difference shown here, especially for $\tau_1$ may be exaggerated due to the low sampling rate used for the samples in water.

\subsection{Magnetic field dependence of the MPL effect}

\begin{figure}
	\centering
	\includegraphics{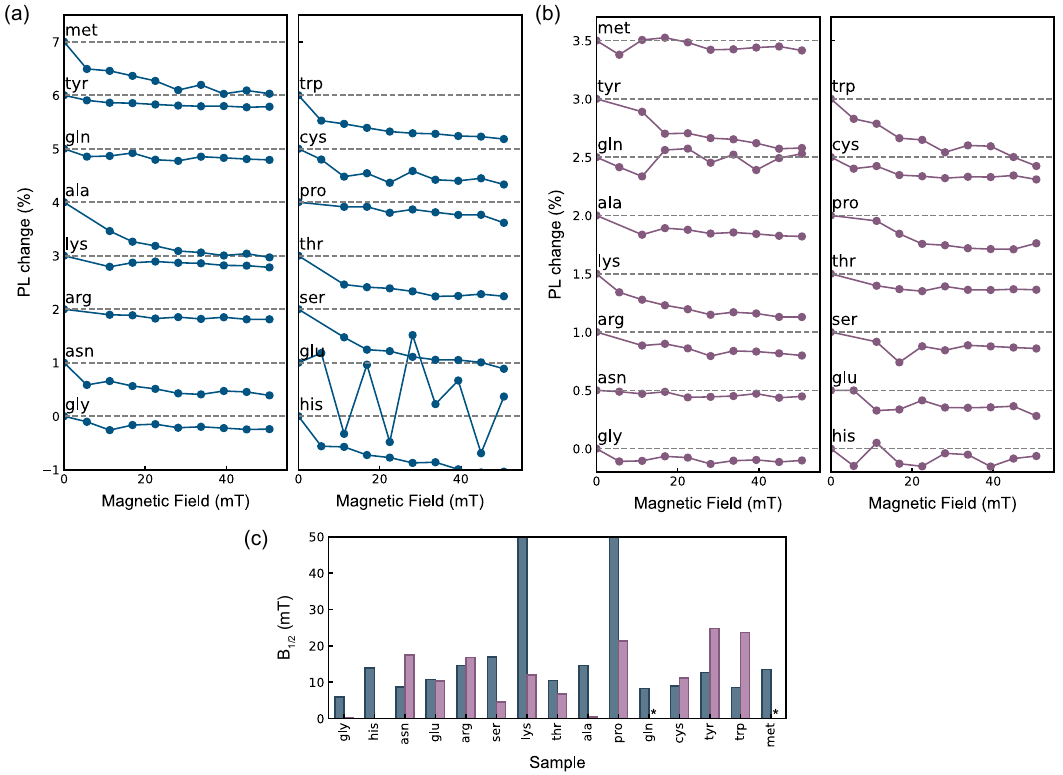}
	\caption{
		\textbf{a},\textbf{b} PL change as a function of applied magnetic field measured for 15 samples both dry (a) and in water (b). 
		\textbf{c} The fitted $B_{1/2}$ field from the data in \textbf{a} and \textbf{b}. An asterisk  indicates that the fit was not reliable, while non-observable bars indicate that the decay occurred well below the first magnetic field value (6 mT). 
	}
	\label{SI_MPL vsB}
\end{figure}

The magnetic field dependence of the MPL effect was measured by switching $B$ on and off while increasing its value between each on-off cycle. The background was removed using the procedure described earlier, i.e. using the off periods to determine the background trend. A $B$ ramp up going from 0 to 46\,mT was immediately followed by a ramp down from 46 to 0\,mT, and the results from the two ramps averaged to increase the signal to noise (no hysteresis was observed). The resulting curves are presented for 15 samples in Fig. \ref{SI_MPL vsB}a,b. As explained in the main text, by fitting the curves with a stretched-exponential curve we can extract the $B_{1/2}$ field at which we obtain half of the saturation contrast, plotted in Fig. \ref{SI_MPL vsB}c.      

\subsection{Laser power dependence of MPL dynamics}

\begin{figure}
	\centering
	\includegraphics[width=\columnwidth]{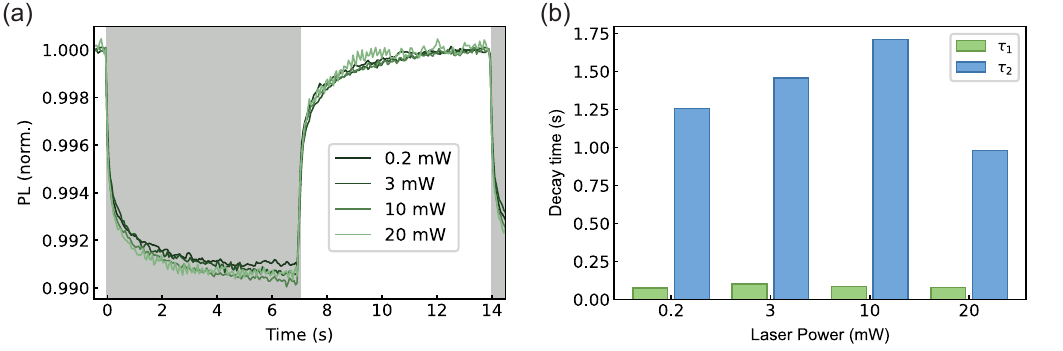}
	\caption{
		(a) The averaged PL trace taken using different laser powers. 
		(b) The extracted decay times, $\tau_1$ and $\tau_2$, from the B-on condition at each laser power. Error bars are not visible. 
	}
	\label{SI_MPLvsLP}
\end{figure}

We tested whether variation in laser intensity could explain the differences in decay times shown in Fig. \ref{SI_Taus_aaCQDs}. To do this we subjected sample \textit{Tyr}-CQD to variations in laser intensity over two orders of magnitude. The decay times $\tau_1$ and $\tau_2$ were extracted from a bi-exponential fit to the data as previously. Fig. \ref{SI_MPLvsLP} shows minimal variation across this range of excitation intensities. From this we conclude that the differences in dynamic behaviour we see across the samples stems from their physical properties as opposed to any potential variations in laser intensity they may have experienced during a measurement.

\subsection{MPL measurements on control samples}

\begin{figure}
	\centering
	\includegraphics[width=\columnwidth]{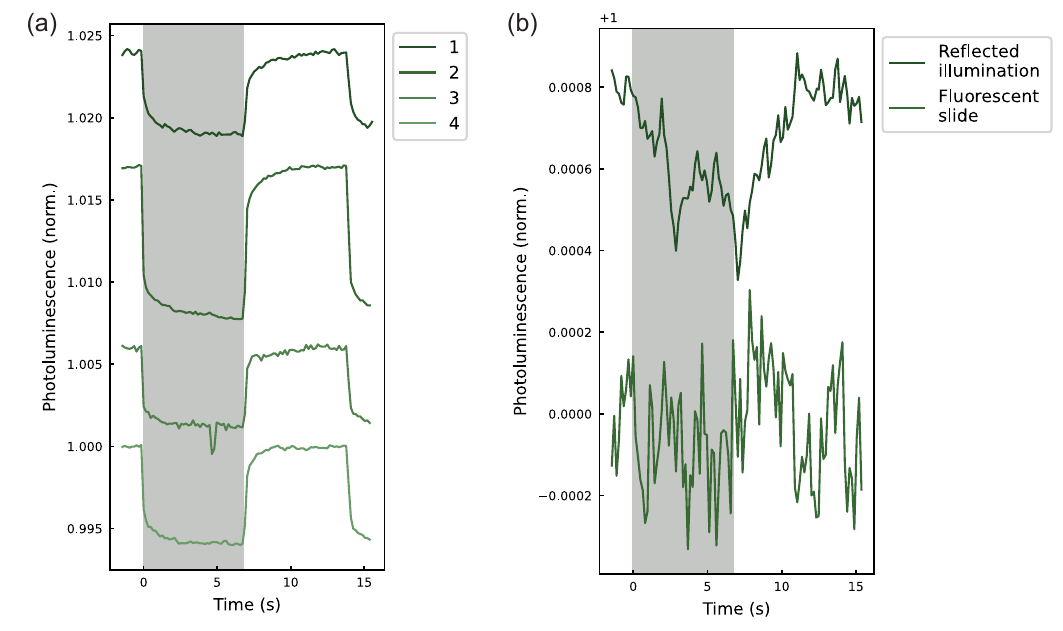}
	\caption{
		(a) PL trace of four dry commercial CQD samples labelled 1-4. 
		(b) MPL data taken from two control measurements. The first is the reflected light off a glass slide, and the second is a fluorescent plastic slide.
	}
	\label{SI_Controls}
\end{figure}

To understand whether the MPL contrast observed in our \textit{aa}-CQDs is unique, we measured four commercial samples purchased from Nanopartz (\#A51-(COLOR)-(WT)). Fig. \ref{SI_Controls}(a) demonstrates that all four of these samples have MPL contrast, indicating that this phenomena is not unique to our samples.

We also tested two control conditions to ensure that the MPL contrast we observed was not the result of a systematic artefact. Fig. \ref{SI_Controls}(b) shows the signal from both the light reflected off a clean coverslip and the fluorescence from a plastic calibration slide. Neither show credible MPL contrast.

\section{Additional ODMR data and methods}

\subsection{Measurement procedure}

\begin{figure}
	\centering
	\includegraphics{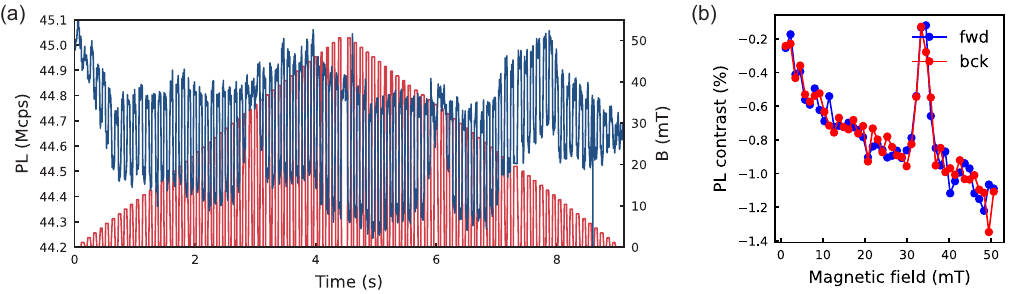}
	\caption{
		\textbf{a} PL trace of dry \textit{gly}-CQDs (blue) under a ramp of magnetic field switched on and off every 5\,s (red) with an applied RF field driving at 980~MHz. A clear suppression of the MPL effect is observed when $B$ meets the ESR condition.  
		\textbf{b} MPL contrast as a function of magnetic field for increasing the field (blue) and decreasing (red). 
	}
	\label{SI_ODMR}
\end{figure}

Electron spin resonance (ESR) was detected optically (i.e.~ODMR) in two different ways. 
Both methods involve driving a radiofrequency (RF) signal at a given or swept frequency that when on resonance with the spin Larmor frequency induces a change in the PL behaviour. 
The first method involved maintaining a fixed RF frequency ($f = 980$\,MHz) and performing a ramp of on-off magnetic pulses as shown in Fig.~\ref{SI_ODMR}. 
As in the no-RF magnetic ramp measurements, as the field is increased the contrast between the on and off PL conditions (i.e. the MPL contrast) increases. 
However, when the field matches the ESR condition, $B=hf/g\mu_B\approx35$\,mT, the MPL contrast is quenched. This is because at resonance the RF mixes the spin states, similar to the $B=0$ condition when hyperfine interactions mix the spin state, hence there is no longer any PL difference between field on and off. A ramp up followed by a ramp down are taken to verify there is no hysteresis [Fig.~\ref{SI_ODMR}b]. 

\begin{figure}
	\centering
	\includegraphics{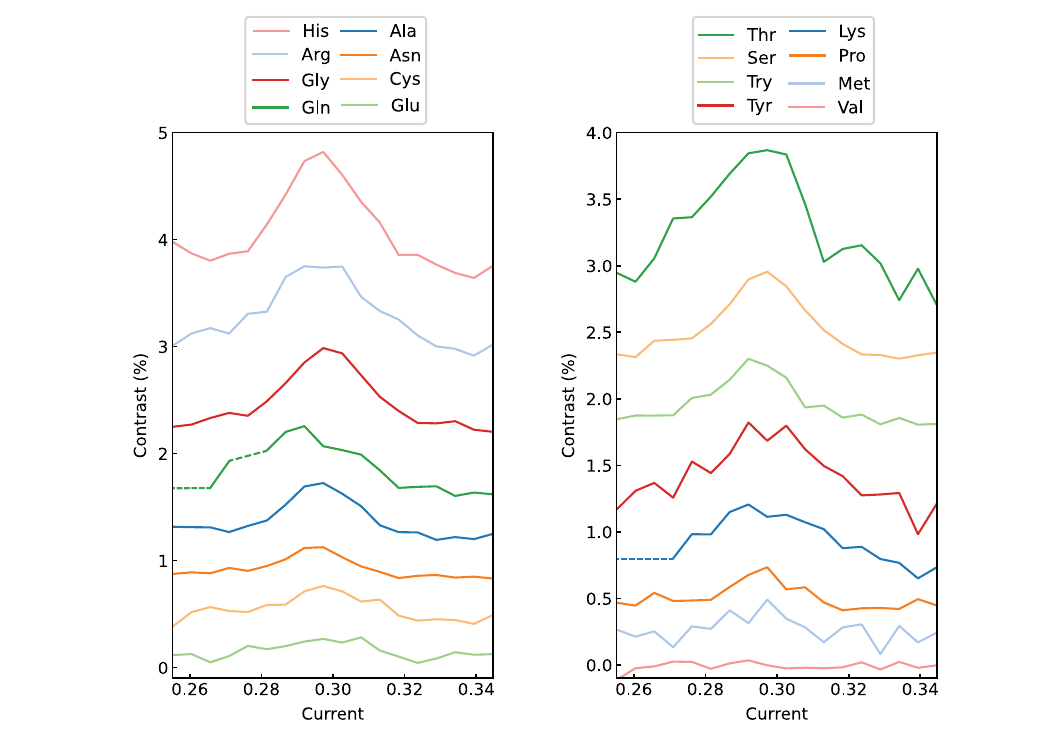}
	\caption{
		ODMR spectra of 17 dry \textit{aa}-CQD samples taken by ramping the magnetic field with a fixed RF driving field at $f = 980$\,MHz. Outliers, replaced by dotted lines, have been removed for clarity. 
	}
	\label{SI_ODMRall}
\end{figure}

The spectra obtained for all samples that showed MPL contrast are shown in Fig. \ref{SI_ODMRall}, which were used to extract the ratio of ODMR and MPL contrasts as explained in the main text. The ESR linewidth is typically around 5\,mT.

The second method to acquire ODMR spectra involved maintaining a fixed magnetic field while sweeping the RF frequency to match the ESR condition, which is the more common procedure for ODMR experiments as it does not require precise control over the magnetic field.
We used this method to measure the $g$-factor of the spins as explained below. 

\subsection{$g$-factor measurement}

\begin{figure}
	\centering
	\includegraphics[width=0.6\linewidth]{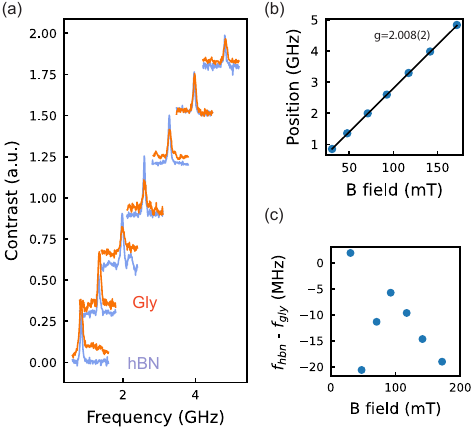}
	\caption{
		\textbf{a} Series of ODMR measurements taken by scanning the RF frequency at different magnetic field values for a hBN sample and dry \textit{gly}-CQD. The second from the bottom data set shows a second lower frequency peak that is a harmonic of the fundamental resonance caused by the amplifier nonlinearity. 
		\textbf{b} Fit of the g-factor to the frequency position of the CQD giving g=2.01(1). 
		\textbf{c} Difference between the frequency of the hBN sample $f_{hBN}$ and the CQD sample $f_{gly}$.
	}
	\label{SI_gfactor}
\end{figure}

We measured the $g$-factor of one of the CQD samples (dry \textit{gly}-CQD) using the well characterised system of spin pairs in hBN as a calibration, which has been previously determined to have a $g$-factor of $g=2.00(1)$ \cite{scholten_multi-species_2024, scholten_optically_2025}. 
For each position of a permanent magnet, we iteratively took an ODMR measurement for both samples (hBN film and \textit{gly}-CQD) by laterally shifting a coverslip with the samples into the optical focus, without moving any other part (magnet, PCB, laser spot position/focus) thus minimising any difference in the magnetic field experience by the two samples. ODMR spectra recorded at different positions of the magnet up to 170\,mT are shown in Fig.~\ref{SI_gfactor}a. 
The peak position for the hBN sample allows us to estimate the $B$ value, which then allows us to plot the ESR frequency versus $B$ relation for \textit{gly}-CQD [Fig.~\ref{SI_gfactor}b]. The $g$-factor for the spins in \textit{gly}-CQD is estimated to be $g=2.01(1)$. There seems to be a measurable (though small) difference between the ESR frequencies of hBN versus \textit{gly}-CQD [Fig.~\ref{SI_gfactor}c] except for two outlying points that experienced RF board resonances that made the fit less reliable. In any case, the difference is close to our measurement uncertainty and so we conclude that the $g$-factor of \textit{gly}-CQD spins is essentially within error of those in hBN and of free electrons.

\subsection{Observation of Rabi oscillations}

By pulsing the laser and the RF field as depicted in Fig. \ref{SI_rabi}a, we can attempt to observe Rabi oscillations as the duration of the RF pulse (at the ESR condition) is increased. The example of dry \textit{gly}-CQD is shown in Fig. \ref{SI_rabi}b (raw PL traces with and without the RF pulse) and the normalised curve is shown in Fig. \ref{SI_rabi}c. One oscillation can be observed, with a short damping time of less than 100\,ns.   

\begin{figure*}
	\centering
	\includegraphics{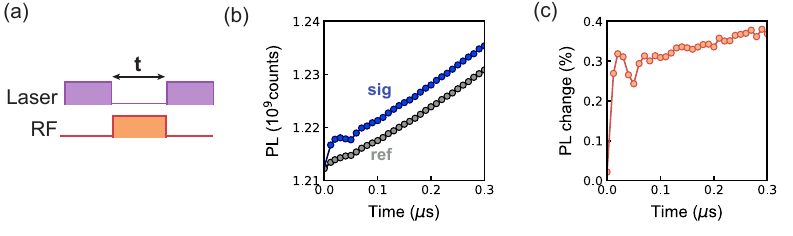}
	\caption{
		\textbf{a} Rabi pulse sequence where the laser illumination and RF driving are interleaved with variable time.
		\textbf{b} Non-normalised Rabi curve with the signal (blue) with RF on and the reference (black) without RF, measured for \textit{gly}-CQD.
		\textbf{c} Normalised Rabi curve. 
	}
	\label{SI_rabi}
\end{figure*}

\section{Changes in PL spectrum with ionic solutions}

\begin{figure*}
	\centering
	\includegraphics{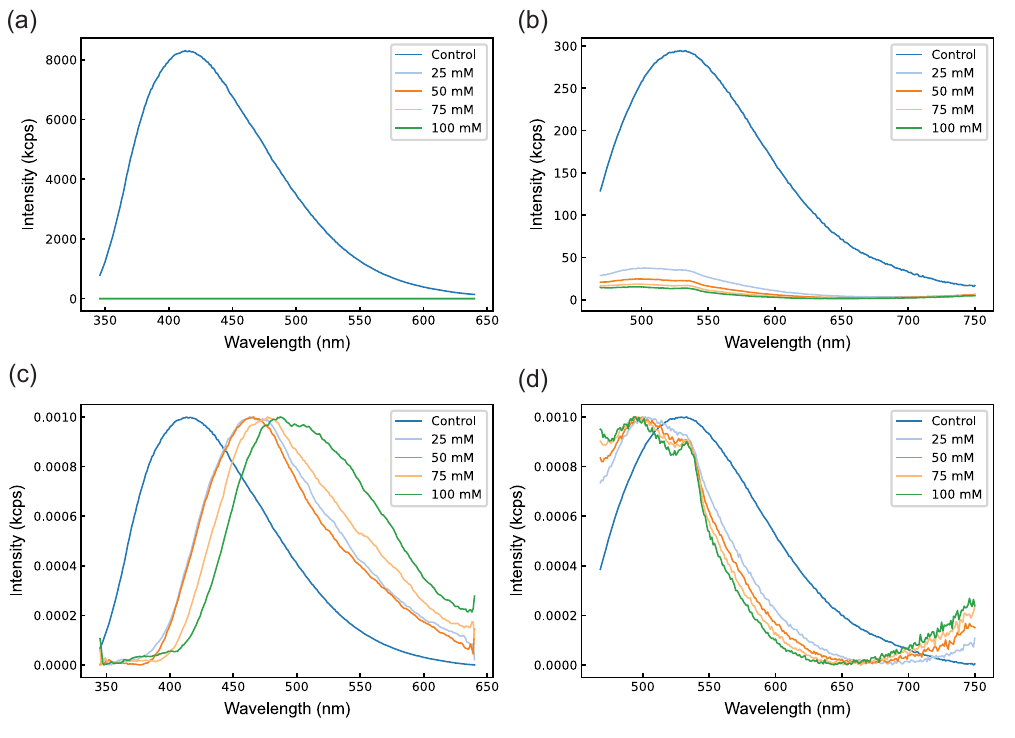}
	\caption{ \textit{Gly}-CQD with varying concentrations of 
		(a) iron(III) chloride and
		(b) copper(II) acetate. Panels (c) and (d) are normalised versions of (a) and (b) respectively, to show shifts in peak wavelength.
	}
	\label{SI_rabi}
\end{figure*}

\section{Demonstration of lock-in imaging using MPL}

Here we demonstrate a potential application of the MPL effect in CQDs for fluorescence imaging. The idea is that if the CQDs constitute the signal of interest (e.g. they could be bound to a molecule one wishes to track in a cell), the background fluorescence (e.g. cell auto-fluorescence or other fluorescent dyes) can be removed by measuring the MPL signal instead of the total PL. Since the MPL signal is measured by continuously modulating the applied magnetic field, this measurement is a form of lock-in detection. This technique could be particularly beneficial in environments where there is naturally a large background fluorescence that is difficult to spectrally remove.   

\begin{figure}
	\centering
	\includegraphics[width=1\linewidth]{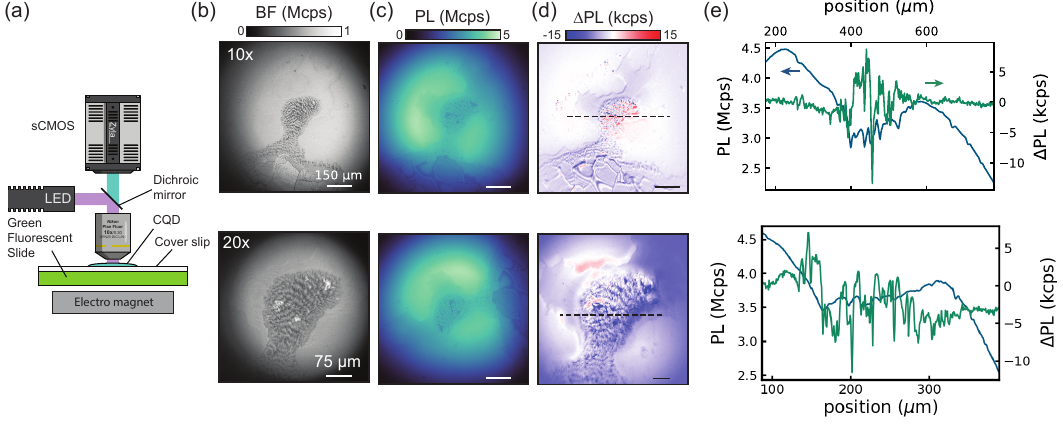}
	\caption{
		\textbf{a} Illustration of the setup for MPL-based lock-in imaging, which includes the introduction of a bright fluorscent slide to simulate a bright background environment. 
		\textbf{b},\textbf{c} Bright-field PL images respectively of deposited \textit{his}-QD for 10x (top panel) and 20x (bottom panel) objectives. 
		\textbf{d} Difference in the PL ($\Delta$PL) between with and without magnetic field.
		\textbf{e} Line cut through in $\Delta$PL as indicated in \textbf{d}
	}
	\label{fig:bck sub}
\end{figure}

To demonstrate the principle of MPL-based lock-in imaging, we placed a coverslip with dried \textit{his}-CQDs on top of a fluorescent green slide (Thorlabs FSK2) giving a large background fluorescence (Fig.~\ref{fig:bck sub}\textbf{a}).
To get a homogeneous illumination, we use LED illumination (365~nm) and a sCMOS camera to image a large area of the sample (Fig.~\ref{fig:bck sub}\textbf{a}). 
In a reflective bright field image (Fig.~\ref{fig:bck sub}\textbf{b}) the \textit{his}-CQDs can be seen as darker regions, while under LED illumination the \textit{his}-CQD PL is difficult to observe (Fig.~\ref{fig:bck sub}\textbf{c}), having significantly less PL than (and even blocking some of) the background. 
However, after magnetic field modulation the absolute difference in the PL between $B$ on and $B$ off show a clear signal where the CQDs are, and even show hard to see features in the bright field image.
This is more evident in the line cut of the PL and $\Delta$PL, where in the PL a small approximately 10\% drop in PL is observed in the thick \textit{his}-CQD region versus several orders of magnitude increase in the difference compared with the background (Fig.~\ref{fig:bck sub}\textbf{e}).

\bibliographystyle{naturemag}
\bibliography{bib.bib}

\end{document}